\documentclass[aps,preprintnumbers,amsmath,amssymb,nofootinbib,superscriptaddress,notitlepage,11pt]{revtex4-1}

\usepackage{epsfig}
\usepackage[utf8]{inputenc}
\usepackage{subfigure}
\usepackage{multirow}
\usepackage[figuresright]{rotating}
\usepackage{graphicx}
\usepackage{color}
\usepackage{xcolor}
\usepackage[normalem]{ulem}
\usepackage{pifont}
\usepackage{booktabs}
\usepackage{longtable,booktabs}
\usepackage{diagbox}
\usepackage{afterpage}
\usepackage{tabularx}
\usepackage{slashed}
\usepackage{placeins}
\usepackage{float}
\usepackage{hyperref}

\newcommand{\itp}{\affiliation{Institute of Theoretical Physics, Chinese Academy of Sciences, Beijing 100190, China}}
\newcommand{\ucas}{\affiliation{School of Physical Sciences, University of Chinese Academy of Sciences, Beijing 100049, China}}

\begin{document}

\title{Effects of strange molecular partners of $P_c$ states in $\gamma p \to K \Sigma$ reactions}
\author{Jian-Cheng Suo}\email{suojiancheng@itp.ac.cn}\itp\ucas
\author{Di Ben}
\email{bendi20@mails.ucas.ac.cn}
\affiliation{Department of Physics and High Energy Physics Center, Tsinghua University, Beijing 100084, China}
\author{Bing-Song Zou}\email{zoubs@mail.tsinghua.edu.cn}\affiliation{Department of Physics and High Energy Physics Center, Tsinghua University, Beijing 100084, China}
\date{\today}

\begin{abstract} \rule{0ex}{3ex}
Our previous studies revealed evidence of the strange molecular partners of $P_c$ states, $N(2080)3/2^-$ and $N(2270)3/2^-$, in the $\gamma p \to K^{*+} \Sigma^0 / K^{*0} \Sigma^+$ and $\gamma p \to \phi p$ reactions. Motivated by the differential cross-section data for $\gamma p \to K^+ \Sigma^0$ from CLAS 2010, which exhibits some bump structures at $W \approx$ 1875, 2080 and 2270 MeV, we extend our previous analysis by investigating the effects of $N(1535)1/2^-$, $N(1875)3/2^-$, $N(2080)1/2^- \&\ 3/2^-$ and $N(2270)1/2^- , 3/2^- \&\ 5/2^-$, as strange partners of $P_c$ molecular states, in the reactions $\gamma p \to K^+ \Sigma^0$ and $\gamma p \to K^0 \Sigma^+$. The theoretical model employed in this study utilizes an effective Lagrangian approach in the tree-level Born approximation. It contains the contributions from $s$-channel with exchanges of $N$, $\Delta$, $N^*$ (including the hadronic molecules with hidden strangeness), and $\Delta^*$; $t$-channel; $u$-channel; and the generalized contact term. The results based on the final fitted parameters are in good agreement with all available experimental data of both cross-sections and polarization observables for $\gamma p \to K^+ \Sigma^0$ and $\gamma p \to K^0 \Sigma^+$. Notably, the $s$-channel exchanges of molecules significantly contribute to the bump structures in cross-sections for $\gamma p \to K \Sigma$ at $W \approx$ 1900, 2080 and 2270 MeV, and show considerable coherence with contributions from $s$-channel exchanges of general resonances to construct the overall structures of cross-sections. More abundant experiments, particularly for the reaction $\gamma p \to K^0 \Sigma^+$, are necessary to further strengthen the constraints on the theoretical models.
\end{abstract}

\maketitle

\section{INTRODUCTION}\label{sec:INTRODUCTION}
The several $P_c$ states observed by the LHCb experiment in 2015 and later~\cite{LHCb:2015yax,LHCb:2019kea} are the most convincing multiquark candidates, prompting significant interest in investigating their nature~\cite{Chen:2016qju,Guo:2017jvc,Liu:2019zoy}. In the hadronic molecular picture, the $P_c(4312)$ can be interpreted as a narrow $\overline{D}\Sigma_c$ bound state with spin-parity $J^P = 1/2^-$, while the $P_c(4440)$ and $P_c(4457)$ can be interpreted as two degenerate narrow $\overline{D}^*\Sigma_c$ bound states with $J^P = 1/2^- \&\ 3/2^-$, respectively~\cite{He:2019rva,Chen:2019asm,Liu:2019zvb}. Moreover, a  $\overline{D}\Sigma_c^*$ bound state with $J^P = 3/2^-$ referred to as $P_c(4380)$, which is different from the broad one reported by LHCb in 2015, and three  $\overline{D}^*\Sigma_c^*$ bound states with $J^P = 1/2^-, 3/2^- \&\ 5/2^-$ are also expected to exist, based on heavy quark spin symmetry~\cite{Du:2019pij,Du:2021fmf,Liu:2019tjn,Yalikun:2021bfm}. The successful interpretation of these hidden-charm $P_c$ states as the hadronic molecules inspires us to investigate their strange partners. 

In the strange sector, S-wave $K\Sigma^*$ molecule $N(1875)3/2^-$, S-wave $K^*\Sigma$ molecules $N(2080)1/2^- \&\ 3/2^-$ and S-wave $K^*\Sigma^*$ molecules $N(2270)1/2^- , 3/2^- \&\ 5/2^-$ are proposed as the strange partners of $P_c$ molecular states~\cite{He:2017aps,Zou:2018uji,Lin:2018kcc,Ben:2023uev,Wu:2023ywu,Ben:2024qeg}. In Refs.~\cite{Lin:2018kcc,Ben:2024qeg}, their decay patterns have been calculated using an effective Lagrangian approach. Notably, in the most recent Review of Particle Physics(RPP)~\cite{ParticleDataGroup:2024cfk}, the two-star $N(2080)$ listed before the 2012 review has been split into two three-star states: $N(1875)$ and $N(2120)$. For consistency with our previous work, we retain the designation $N(2080)3/2^-$ for the possible $K^*\Sigma$ molecule, which is not necessarily identified with the $N(2120)$ resonance in RPP~\cite{ParticleDataGroup:2024cfk}. Furthermore, the contentious state $N(1535)1/2^-$ can also be interpreted as a bound state of $K\Lambda,K\Sigma$ within the molecular picture~\cite{Zou:2018uji,Kaiser:1995cy,Bruns:2010sv,Li:2023pjx,Molina:2023jov}.

We have conducted several studies to investigate the effects of these hidden-strange molecules in photoproduction reactions. In Refs.~\cite{Ben:2023uev,Wu:2023ywu}, the $N(2080)3/2^-$ and $N(2270)3/2^-$ are introduced in s-channel as the primary contributors to the $\gamma p \to K^{*+} \Sigma^0 / K^{*0} \Sigma^+$ and $\gamma p \to \phi p$ reactions. The theoretical models constructed based on this fit well with the available experimental data for these reactions. Following this, we observe that the differential cross-section data for $\gamma p \to K^+ \Sigma^0$ from CLAS 2010~\cite{CLAS:2010aen} exhibits bump structures near the center-of-mass energies $W$ = 1875, 2080 and 2270 MeV, as shown in Fig.~\ref{dsigmap}, corresponding to the Breit-Wigner masses of $N(1875)3/2^-$, $N(2080)1/2^- \&\ 3/2^-$ and $N(2270)1/2^-, 3/2^- \&\ 5/2^-$. Additionally, the $K\Sigma$ channel is essential in the molecular picture of $N(1535)1/2^-$~\cite{Molina:2023jov}. These prompt us to focus on the $\gamma p \to K^+ \Sigma^0$ and $\gamma p \to K^0 \Sigma^+$ reactions to test the effects of these seven hidden-strange molecules mentioned above.

The $K \Sigma$ photoproduction reactions have garnered significant attention both experimentally and theoretically over the past few years, contributing to the study of the light baryon resonance spectrum. In the experimental aspect, various collaborations such as CLAS, SAPHIR, LEPS have contributed large and diverse sets of experimental data on both cross-sections and polarization observables for the reaction $\gamma p \to K^+ \Sigma^0$~\cite{Glander:2003jw,Kohri:2006yx,CLAS:2005lui,CLAS:2010aen,Schmieden:2014oba,LEPS:2017pzl,Jude:2020byj,Lleres:2007tx,CLAS:2016wrl,LEPS:2017pzl,CLAS:2006pde}. With the exception of some older measurements, these data generally show no significant discrepancies. For the reaction $\gamma p \to K^0 \Sigma^+$, several collaborations such as SAPHIR, CBELSA, A2 have also provided the experimental data~\cite{Lawall:2005np,CBELSATAPS:2007oqn,A2:2013cqk,CBELSATAPS:2011gly,CLAS:2013owj,CLAS:2024bzi}, including the latest data of polarization observables from the CLAS Collaboration~\cite{CLAS:2024bzi}. However, in comparison to $\gamma p \to K^+ \Sigma^0$, the amount of experimental data available for $\gamma p \to K^0 \Sigma^+$ remains relatively sparse.

Many theoretical works have already been devoted to analyzing the data for $K\Sigma$ photoproduction, based on the effective Lagrangian approaches, isobar models, Regge-plus-resonance models, and so on~\cite{Wei:2022nqp,Wei:2023gvh,Ronchen:2022hqk,Mart:2019fau,Clymton:2021wof,Maxwell:2016hdx,Sarantsev:2005tg,Steininger:1996xw,Mai:2009ce,Kaiser:1996js,Borasoy:2007ku,Golli:2016dlj,Luthfiyah:2021yqe,Egorov:2020ghz,Lee:1999kd,Tiator:2018pjq,Corthals:2006nz}. In Refs.~\cite{Mart:2019fau,Clymton:2021wof} and Ref.~\cite{Ronchen:2022hqk}, photoproduction data for $K\Sigma$ have been simultaneously analyzed and effectively described using an isobar model and the Jülich-Bonn dynamical coupled-channel approach, respectively. And the work in Refs.~\cite{Wei:2022nqp,Wei:2023gvh} provides a comprehensive analysis of the available data for $\gamma n \to K^+ \Sigma^-$ and $\gamma n \to K^0 \Sigma^0$ reactions, based on an effective Lagrangian approach in the tree-level Born approximation. 

In this work, we employ the methodology used in Refs.~\cite{Wei:2022nqp,Wei:2023gvh} to simultaneously analyze data for $\gamma p \to K^+ \Sigma^0$ and $\gamma p \to K^0 \Sigma^+$ reactions. Our theoretical model incorporates contributions from $s$-channel exchanges of $N$, $\Delta$, $N^*$(including the hadronic molecules with hidden strangeness), and $\Delta^*$; $t$-channel exchanges of $K$, $K^*(892)$, and $K_1(1270)$; $u$-channel exchange of $\Sigma$; and the generalized contact term. We utilize this model to investigate the reaction mechanisms and test the effects of hidden-strange molecules in $\gamma p \to K \Sigma$ reactions.

The article is organized as follows. Sec.~\ref{sec:FORMALISM} outlines the framework of our theoretical model. Sec.~\ref{sec:FITTING SETTINGS} details our fitting settings. Sec.~\ref{sec:RESULTS AND DISCUSSIONS} presents the results of our theoretical model along with some discussions. Finally, Sec.~\ref{sec:SUMMARY AND CONCLUSION} provides the summary and conclusions.

\section{FORMALISM}\label{sec:FORMALISM}
\begin{figure}[H]
\centering
\subfigure[$s$-channel]
{
\includegraphics[scale=0.5]{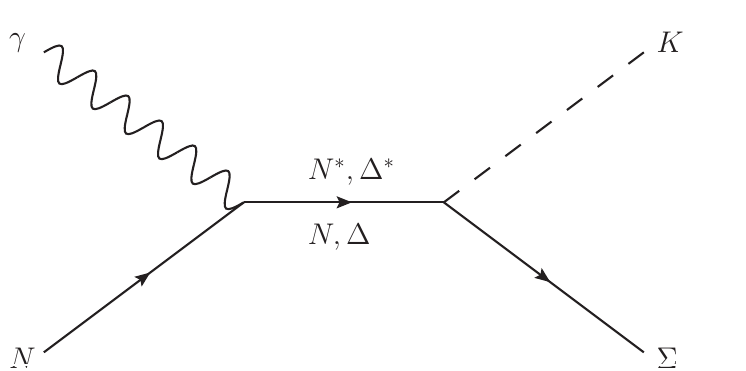} 
\label{s-channel} 
}
\quad
\subfigure[$t$-channel]
{
\includegraphics[scale=0.5]{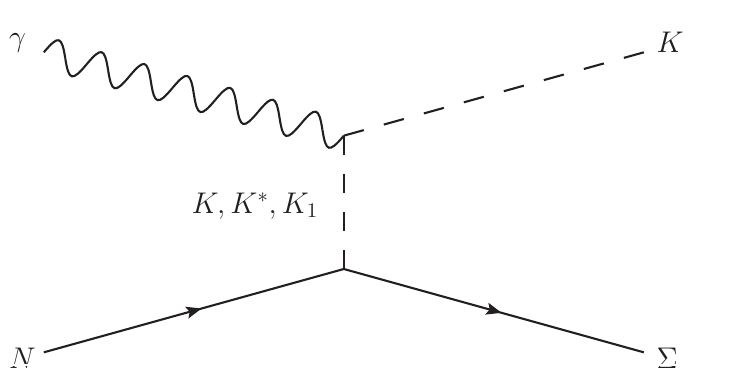} 
\label{t-channel} 
}

\subfigure[$u$-channel]
{
\includegraphics[scale=0.5]{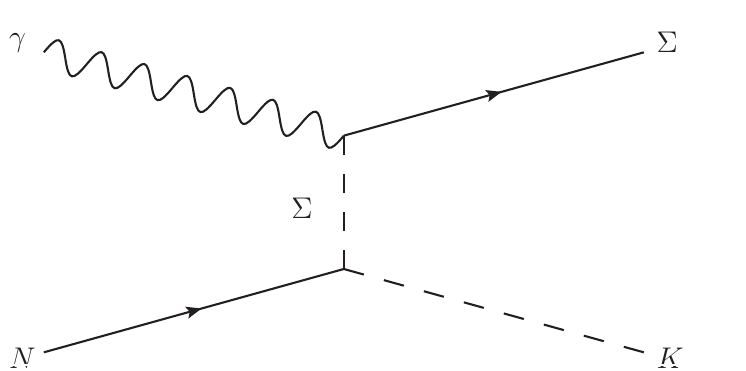} 
\label{u-channel} 
}
\quad
\subfigure[Interaction current]
{
\includegraphics[scale=0.5]{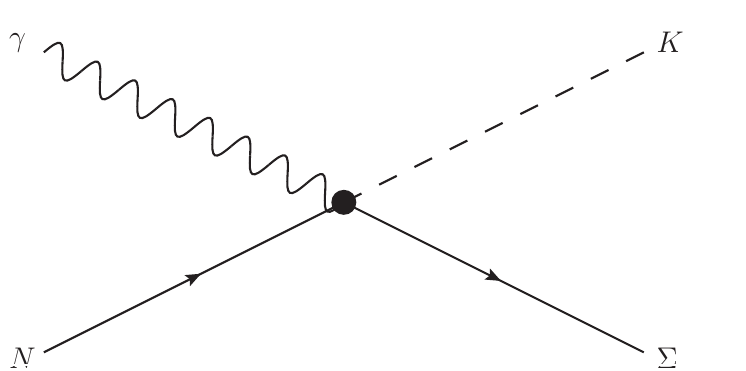} 
\label{interaction} 
}

\caption{Generic structure of the amplitude for $\gamma N \to K \Sigma$. Time proceeds from left to right.}
\label{feynman}
\end{figure}

As shown in Fig.~\ref{feynman}, the gauge-invariant amplitude of $K\Sigma$ photoproduction reactions in the tree-level effective Lagrangian approach can be expressed as~\cite{Wei:2022nqp,Wei:2023gvh}
\begin{align}
  M = M_s + M_t + M_u + M_{int},
\label{Eq1}
\end{align} 
where the terms $M_s$, $M_t$, $M_u$ and $M_{int}$ stand for the amplitudes calculated from the $s$-channel mechanism, $t$-channel mechanism, $u$-channel mechanism and the interaction current, respectively. The first three terms in Eq.~(\ref{Eq1}) can be directly computed from the effective Lagrangians (with additional phase factors for molecules), progagators and form factors, which will be elaborated in the subsequent part of this section. The specifics of the last term will also be detailed in Sec.~\ref{sec:Interaction current}. Furthermore, the mathematical definitions of polarization observables analyzed in this work are provided in Sec.~\ref{sec:Definitions of Polarization Observables}.

\subsection{$s$-channel}\label{sec:s-channel}
Fig.~\ref{s-channel} presents the $s$-channel with exchanges of $N$, $\Delta$, $N^*$, and $\Delta^*$. 
First, to investigate the effects of hidden-strange molecular states in $K \Sigma$ photoproduction, we introduce the seven molecules: $N(1535)1/2^-$, $N(1875)3/2^-$, $N(2080)1/2^- \&\ 3/2^-$ and $N(2270)1/2^-, 3/2^- \&\ 5/2^-$. Second, in Ref.~\cite{Wei:2023gvh}, contributions from the $N(1710)1/2^+$, $N(1880)1/2^+$, $N(1900)3/2^+$, $N(1895)1/2^-$, $N(2060)5/2^-$, $\Delta(1910)1/2^+$ and $\Delta(1920)3/2^+$ resonances have been taken into account to reproduce the available data for both $\gamma n \to K^0 \Sigma^0$ and $\gamma n \to K^+ \Sigma^-$ reactions. Apart from the $N(1710)1/2^+$ which is marked as "seen" in its decay branching ratio to the $K \Sigma$ channel, all the other considered resonances have sizable branching ratios in RPP~\cite{ParticleDataGroup:2024cfk}. In this work, we disregard the $N(1895)1/2^-$ and $N(2060)5/2^-$ for the final fit. We performed preliminary fits including these two states and found their contributions to be negligible, yielding no significant improvement to the overall $\chi^2$. A possible reason is that their contribution is somewhat diluted by the presence of nearby molecular states with similar masses. Therefore, to avoid over-parameterization, these two states were not included in our final model configuration. We retain the other five resonances: $N(1710)1/2^+$, $N(1880)1/2^+$, $N(1900)3/2^+$, $\Delta(1910)1/2^+$ and $\Delta(1920)3/2^+$. Third, to achieve satisfactory numerical results, we refer to the analyses in Refs.~\cite{Mart:2019fau,Clymton:2021wof} and add seven additional resonances that may contribute significantly: $N(1675)5/2^-$, $N(1720)3/2^+$, $\Delta(1600)3/2^+$, $\Delta(1700)3/2^-$, $\Delta(1900)1/2^-$, $\Delta(1930)5/2^-$ and $\Delta(1940)3/2^-$. In summary, besides the ground states $N$ and $\Delta$, there are seven molecules, five general $N^*$ resonances and seven $\Delta^*$ resonances considered in $s$-channel of our theoretical model, which are listed in Tab.~\ref{Table3}.

The effective Lagrangians for ground states $N$ and $\Delta$ are presented in Sec.~\ref{sec:tu-channel}, along with the Lagrangians of other background terms. The general forms of the effective Lagrangians for the electromagnetic and hadronic interactions of s-channel resonances ($R$) are given below. For brevity, we define $\Gamma^{(+)}=\gamma_{5}$ and $\Gamma^{(-)}=1$, and the photon field-strength tensor is $F^{\mu\nu}=\partial^{\mu}A^{\nu}-\partial^{\nu}A^{\mu}$. These forms are used for both the general resonances and the molecular candidates, corresponding to their respective spin-parity ($J^P$)~\cite{Wei:2022nqp}.
\paragraph{Electromagnetic Couplings:}
\allowdisplaybreaks
\begin{align}
&  \mathcal{L}_{RN\gamma}^{1/2\pm} = e\frac{g_{RN\gamma}^{(1)}}{2M_{N}}\overline{R}\Gamma^{(\mp)}\sigma_{\mu\nu}(\partial^{\nu}A^{\mu})N+\text{H.c.}, \\
&  \mathcal{L}_{RN\gamma}^{3/2\pm} = -ie\frac{g_{RN\gamma}^{(1)}}{2M_{N}}\overline{R}_{\mu}\gamma_{\nu}\Gamma^{(\pm)}F^{\mu\nu}N +e\frac{g_{RN\gamma}^{(2)}}{(2M_{N})^{2}}\overline{R}_{\mu}\Gamma^{(\pm)}F^{\mu\nu}\partial_{\nu}N+\text{H.c.}, \\
&  \mathcal{L}_{RN\gamma}^{5/2\pm} = e\frac{g_{RN\gamma}^{(1)}}{(2M_{N})^{2}}\overline{R}_{\mu\alpha}\gamma_{\nu}\Gamma^{(\mp)}(\partial^{\alpha}F^{\mu\nu})N \pm ie\frac{g_{RN\gamma}^{(2)}}{(2M_{N})^{3}}\overline{R}_{\mu\alpha}\Gamma^{(\mp)}(\partial^{\alpha}F^{\mu\nu})\partial_{\nu}N + \text{H.c.}.
\end{align}
\paragraph{Hadronic Couplings to K$\Sigma$:}
\allowdisplaybreaks
\begin{align}
&  \mathcal{L}_{R\Sigma K}^{1/2\pm} = -g_{R\Sigma K}\overline{\Sigma}\Gamma^{(\pm)}\left[\left(i\lambda+\frac{1-\lambda}{M_{R}\pm M_{\Sigma}}\slashed{\partial}\right)K\right]R + \text{H.c.}, \\
&  \mathcal{L}_{R\Sigma K}^{3/2\pm} = \pm\frac{g_{R\Sigma K}}{M_{K}}\overline{\Sigma}\Gamma^{(\mp)}(\partial^{\mu}K)R_{\mu}+\text{H.c.}, \\
&  \mathcal{L}_{R\Sigma K}^{5/2\pm} = i\frac{g_{R\Sigma K}}{M_{K}^{2}}\overline{\Sigma}\Gamma^{(\pm)}(\partial^{\mu}\partial^{\nu}K)R_{\mu\nu}+\text{H.c.}.
\end{align}
Here, $e$ is the elementary charge unit. $M_i$ are the masses of the corresponding particles. The parameter $\lambda$ mixes the pseudoscalar ($\lambda=1$) and pseudovector ($\lambda=0$) couplings; we use the pure pseudovector coupling, i.e., $\lambda=0$, in our calculations and set $g_{\gamma NR}g_{R\Sigma K}$ as fitting parameters since the reaction amplitudes are only sensitive to the products of electromagnetic and hadronic coupling constants.

\begin{figure}[H]
\centering

\includegraphics[scale=0.7]{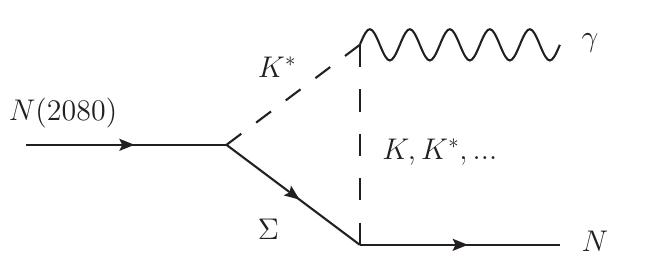} \\
\includegraphics[scale=0.7]{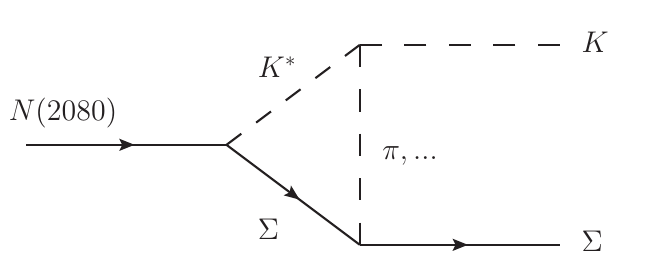} 

\caption{Electromagnetic and hadronic couplings of $N(2080)3/2^-$ as $K^* \Sigma$ molecule.}
\label{N2080}
\end{figure}

Additionally, we provide a brief explanation of the treatment of the molecules in the $s$-channel. Take $N(2080)3/2^-$ as an example, which is assumed to be a pure $S$-wave molecular state of $K^*$ and $\Sigma$. In the hadronic molecular picture, both the electromagnetic and hadronic couplings of it are dedicated by the loop diagrams illustrated in Fig.~\ref{N2080}~\cite{Lin:2018kcc}. Here for simplicity, we just follow Ref.~\cite{Ben:2023uev} to calculate the tree-level approximation by using the effective Lagrangians of $N^*$ with spin-parity $J^P = 3/2^-$ for $N(2080)3/2^-$:
\begin{align}
&  \mathcal{L}_{\gamma N R}^{3/2^-} = - i e\frac{g_{\gamma N R}^{(1)}}{2 M_N} \overline{R}_\mu \gamma_\nu F^{\mu \nu} N + e\frac{g_{\gamma N R}^{(2)}}{(2 M_N)^2} \overline{R}_\mu F^{\mu \nu} \partial_\nu N + H.c. ,      \\
&  \mathcal{L}_{K \Sigma R}^{3/2^-} = - \frac{g_{K \Sigma R}}{M_K} \overline{\Sigma} \gamma_5 \left( \partial_\mu K \right) R^\mu + H.c. .
\label{N(2080)}
\end{align} 
In addition, we attach a phase factor Exp[$i\phi_R$] in front of the tree-level amplitude, to partially mimic the loop contributions as illustrated in Fig.~\ref{N2080}. Similarly, all the other hidden-strange molecules are treated in the same manner, and the Lagrangians introduced for them are presented above. The masses of $N(1535)1/2^-$, $N(1875)3/2^-$, $N(2080)1/2^- \&\ 3/2^-$, $N(2270)1/2^- , 3/2^- \&\ 5/2^-$ are taken as 1535, 1875, 2080 and 2270 MeV, respectively. Furthermore, the width $\Gamma_R$ and products of coupling constants $g_{\gamma N R}^{(1)}g_{K \Sigma R}$, $g_{\gamma N R}^{(2)}g_{K \Sigma R}$—which depend on the choice of cutoff parameters in Refs.~\cite{Lin:2018kcc,Ben:2024qeg}—along with the phase $\phi_R$ of molecules, are treated as fit parameters.

Here, we would like to emphasize the thought behind our approach to handling the molecular states. Our primary goal of this work was to search for resonance signals in the experimental data that match the specific masses and spin-parities predicted for these molecular candidates. From a purely formal perspective, the primary impact of interpreting these states as molecules versus ordinary resonances is the introduction of the phase factor, which allows for different and potentially enhanced interference patterns. Most importantly, by fitting with such a model, we can explore whether resonance signals consistent with these molecules might exist in the experimental data.

\subsection{$t$- and $u$-channel}\label{sec:tu-channel}
Fig.~\ref{t-channel} illustrates the $t$-channel, which includes exchanges of $K$ and $K^*(892)$ as considered in Refs.~\cite{Wei:2022nqp,Wei:2023gvh}, along with the $K_1(1270)$, which may also contribute. Fig.~\ref{u-channel} depicts the $u$-channel with only the exchange of $\Sigma$. 

It is important to clarify the selection principle for the background contributions in our model. The primary focus of this work is to investigate the effects of the $s$-channel resonances, particularly the hadronic molecular candidates. Therefore, we adopt the principle of employing a minimal yet effective set of background terms. This approach helps to reduce the number of free parameters---many of which would be poorly constrained by existing data---and provides a cleaner background against which the effects of the $s$-channel states can be more clearly discerned. As demonstrated by the excellent fit quality presented in Sec.~\ref{sec:RESULTS AND DISCUSSIONS}, the current selection of background terms is sufficient to achieve a good description of the extensive experimental data. We have also estimated the contribution of the $u$-channel $\Sigma(1385)$ exchange based on the fitted parameter values from Ref.~\cite{Wang:2018vlv}, and found it to be negligibly small. Similarly, Ref.~\cite{Maxwell:2016hdx} notes that adding more resonances in the $u$-channel has not materially improved their result. So the inclusion of more states would likely offer only limited improvement at the cost of reduced stability and clarity in the fit. Based on these considerations, we have not included additional higher-mass exchanges in the background.

The Lagrangians associated with the non-resonant background terms are detailed below, including the Lagrangians for $K_1(1270)$~\cite{Wei:2022nqp,Maxwell:2016hdx}:
\allowdisplaybreaks
\begin{align}
&  \mathcal{L}_{\gamma KK} = ie[K^{+}(\partial_{\mu}K^{-})-K^{-}(\partial_{\mu}K^{+})]A^{\mu}, \\
&  \mathcal{L}_{\Sigma NK} = -g_{\Sigma NK}\overline{\Sigma}\gamma_{5}\left[\left(i\lambda+\frac{1-\lambda}{2M_{N}}\slashed{\partial}\right)K\right]N + \text{H.c.}, \label{eq:11} \\
&  \mathcal{L}_{\gamma KK^{*}} = e\frac{g_{\gamma KK^{*}}}{M_{K}}\epsilon^{\alpha\mu\lambda\nu}(\partial_{\alpha}A_{\mu})(\partial_{\lambda}K)K_{\nu}^{*}, \\
&  \mathcal{L}_{\Sigma NK^{*}} = -g_{\Sigma NK^{*}}\overline{\Sigma}\left[\left(\gamma^{\mu}-\frac{\kappa_{\Sigma NK^{*}}}{2M_{N}}\sigma^{\mu\nu}\partial_{\nu}\right)K_{\mu}^{*}\right]N + \text{H.c.}, \\
& \mathcal{L}_{\gamma K K_1} = - e\frac{g_{\gamma K K_1}}{M_K} \left( \left( \partial_\mu A^\nu \right) K \left( \partial_\nu K_{1}^\mu \right) - \left( \partial^\nu A_\mu \right) K \left( \partial_\nu K_{1}^\mu \right) \right) , \\
& \mathcal{L}_{\Sigma N K_1} = -\overline{\Sigma} \left[ \left(g_{\Sigma N K_1}^{(1)} \gamma^\mu - \frac{g_{\Sigma N K_1}^{(2)}}{2M_N} \sigma^{\mu \nu} \partial_\nu \right) K_{1 \mu} \gamma_5 \right] N + \text{H.c.}, \\
&  \mathcal{L}_{\Sigma\Sigma\gamma} = -e\overline{\Sigma}\left[\left(\hat{e}\gamma^{\mu}-\frac{\hat{\kappa}_{\Sigma}}{2M_{N}}\sigma^{\mu\nu}\partial_{\nu}\right)A_{\mu}\right]\Sigma, \\
&  \mathcal{L}_{\gamma NN} = -e\overline{N}\left[\left(\hat{e}\gamma^{\mu}-\frac{\hat{\kappa}_{N}}{2M_{N}}\sigma^{\mu\nu}\partial_{\nu}\right)A_{\mu}\right]N, \\
&  \mathcal{L}_{\Delta\Sigma K} = \frac{g_{\Delta\Sigma K}}{M_{K}}\overline{\Sigma}(\partial^{\mu}K)\Delta_{\mu}+\text{H.c.}, \\
&  \mathcal{L}_{\Delta N\gamma} = -ie\frac{g_{\Delta N\gamma}^{(1)}}{2M_{N}}\overline{\Delta}_{\mu}\gamma_{\nu}\gamma_{5}F^{\mu\nu}N +e\frac{g_{\Delta N\gamma}^{(2)}}{(2M_{N})^{2}}\overline{\Delta}_{\mu}\gamma_{5}F^{\mu\nu}\partial_{\nu}N+\text{H.c.}, \\
&  \mathcal{L}_{\Sigma NK\gamma} = ig_{\Sigma NK}\frac{1-\lambda}{2M_{N}}\overline{\Sigma}\gamma_{5}\gamma^{\mu}A_{\mu}\hat{Q}_{K}K\tau N. \label{eq:18}
\end{align}
Here, $e$ and $\lambda$ is the same as in Sec.~\ref{sec:s-channel}. Moreover, $\hat{e}$ is the charge operator acting on $N$ or $\Sigma$ field. $\hat{Q}_{K}$ is the charge operator acting on the $K$ field. The anomalous magnetic moment terms are defined as $\hat{\kappa}_{N} \equiv \kappa_{p}\hat{e}+\kappa_{n}(1-\hat{e})$, with the proton and neutron anomalous magnetic moments being $\kappa_{p}=1.793$ and $\kappa_{n}=-1.913$, respectively. Similarly, $\hat{\kappa}_{\Sigma} \equiv \kappa_{\Sigma^{+}}(1+\hat{e})/2+\kappa_{\Sigma^{-}}(1-\hat{e})/2$, with $\kappa_{\Sigma^{+}}=1.458$ and $\kappa_{\Sigma^{-}}=-0.16$. The masses of the corresponding particles are denoted as $M_i$. The Levi-Civita tensor $\epsilon^{\alpha\mu\lambda\nu}$ is defined with the convention $\epsilon^{0123}=+1$.

The coupling constants for the background terms are determined as follows. The radiative coupling constants are $g_{\gamma K^{\pm}K^{*\pm}}=0.413$~\cite{Wei:2022nqp} and $g_{\gamma K^{0}K^{*0}}=-0.631$~\cite{Wang:2018vlv}. The photon-nucleon-delta couplings are determined from the helicity amplitudes of $\Delta\to N\gamma$, giving $g_{\Delta N\gamma}^{(1)}=-4.18$ and $g_{\Delta N\gamma}^{(2)}=4.327$. The strong interaction couplings are determined from flavor SU(3) symmetry, yielding $g_{\Sigma NK}=2.692$, $g_{\Sigma NK^{*}}=-4.26$, $\kappa_{\Sigma NK^{*}}=-2.33$, and $g_{\Delta\Sigma K}=7.89$~\cite{Wei:2022nqp}. Additionally, the $g_{\gamma K K_1}g_{\Sigma N K_1}^{(1)}$ and $g_{\gamma K K_1}g_{\Sigma N K_1}^{(2)}$ are treated as fit parameters.

\subsection{Interaction Current}\label{sec:Interaction current}
Fig.~\ref{interaction} presents the interaction current, which is modeled by a generalized contact current. To ensure the gauge invariance of the full photoproduction amplitude given in Eq.~(\ref{Eq1}), the interaction current $M_{int}$ is modeled by a generalized contact current as Ref.~\cite{Wei:2023gvh}
\begin{align}
  M_{int}^{\mu} = \Gamma_{\Sigma NK}(q)C^{\mu} + M_{\text{KR}}^{\mu}f_{t},
\end{align}
where $\Gamma_{\Sigma NK}(q)$ is the vertex function for the $\Sigma NK$ interaction derived from Eq.~(\ref{eq:11}), and $f_t$ is the form factor for the t-channel $K$ exchange given in Eq.~(\ref{eq:ff_meson_app}). The traditional Kroll-Ruderman term $M_{\text{KR}}^{\mu}$ is derived from the Lagrangian in Eq.~(\ref{eq:18}). The auxiliary current $C^{\mu}$ for the $\gamma p \to K\Sigma$ reaction, which is introduced to ensure that the full photoproduction amplitude satisfies the generalized Ward-Takahashi identity and thus is fully gauge invariant, is choosen as
\begin{align}
    C^{\mu} = -Q_{K}\tau\frac{f_{t}-\hat{F}}{t-q^{2}}(2q-k)^{\mu} - Q_{\Sigma}\tau\frac{f_{u}-\hat{F}}{u-p'^{2}}(2p'-k)^{\mu} - \tau Q_{N}\frac{f_{s}-\hat{F}}{s-p^2}(2p+k)^{\mu},
\end{align}
with the function $\hat{F}$ defined as
\begin{align}
    \hat{F} = 1-\hat{h}(1-\delta_{t}f_{t})(1-\delta_{u}f_{u})(1-\delta_{s}f_{s}).
\end{align}
Here, $k, p, q, p'$ are the four-momenta of the photon $\gamma$, initial nucleon $N$, outgoing $K$, and outgoing $\Sigma$, respectively; $Q_i$ are the electric charges; $\tau$ is the corresponding isospin factor; $f_x$ ($x=s,t,u$) are the form factors for the respective channels given in Sec.~\ref{sec:Phenomenological Form Factors}; the constant $\delta_i = 1$ for non-zero charges $Q_i$, and $\delta_i = 0$ for zero charges $Q_i$; and we set $\hat{h} = 1$ for simplicity.

\subsection{Propagators}\label{sec:Propagators}
The propagators for the exchanged particles are given below. For the $s$-channel nucleons and resonances (spin-1/2, 3/2, and 5/2), we use the following forms~\cite{Wei:2022nqp}:
\begin{align}
&  S_{1/2}(p) = \frac{i(\slashed{p} + M_R)}{p^2 - M_R^2 + iM_R\Gamma_R}, \\
&  S_{3/2}^{\mu\nu}(p) = \frac{i(\slashed{p} + M_R)}{p^2 - M_R^2 + iM_R\Gamma_R} \left( \tilde{g}^{\mu\nu} + \frac{1}{3}\tilde{\gamma}^\mu \tilde{\gamma}^\nu \right), \\
&  S_{5/2}^{\mu\nu,\alpha\beta}(p) = \frac{i(\slashed{p} + M_R)}{p^2 - M_R^2 + iM_R\Gamma_R} \left[ \frac{1}{2}(\tilde{g}^{\mu\alpha}\tilde{g}^{\nu\beta} + \tilde{g}^{\mu\beta}\tilde{g}^{\nu\alpha}) - \frac{1}{5}\tilde{g}^{\mu\nu}\tilde{g}^{\alpha\beta} \right. \nonumber \\
& \hspace{5.5cm} \left. + \frac{1}{10}(\tilde{g}^{\mu\alpha}\tilde{\gamma}^{\nu}\tilde{\gamma}^{\beta} + \tilde{g}^{\mu\beta}\tilde{\gamma}^{\nu}\tilde{\gamma}^{\alpha} + \tilde{g}^{\nu\alpha}\tilde{\gamma}^{\mu}\tilde{\gamma}^{\beta} + \tilde{g}^{\nu\beta}\tilde{\gamma}^{\mu}\tilde{\gamma}^{\alpha}) \right],
\end{align}
with the operators $\tilde{g}^{\mu\nu}$ and $\tilde{\gamma}^{\mu}$ being defined as
\begin{align}
& \tilde{g}^{\mu\nu} = -g^{\mu\nu}+\frac{p^{\mu}p^{\nu}}{M_{R}^{2}}, \\
& \tilde{\gamma}^{\mu} = -\gamma^{\mu}+\frac{p^{\mu}\slashed{p}}{M_{R}^{2}},
\end{align}
where $p$ is the four-momentum, $M_R$ and $\Gamma_R$ are the mass and width of the resonance. For the t-channel meson exchanges, the propagators are given as follows. For the pseudoscalar $K$ meson (spin-0), the propagator is:
\begin{align}
&  S_K(p) = \frac{i}{p^2 - M_K^2}.
\end{align}
For the spin-1 vector meson ($K^*(892)$) and axial-vector meson ($K_1(1270)$), the propagators share the same mathematical form~\cite{Maxwell:2012zz}:
\begin{align}
&  S_{V}^{\mu\nu}(p) = \frac{-i}{p^2 - M_{V}^2} \left( g^{\mu\nu} - \frac{q^\mu q^\nu}{M_{V}^2} \right),
\end{align}
where the subscript $V$ represents the $K^*$ or $K_1$ meson, with the corresponding mass $M_V$ used for each. And no width is included, as the intermediate energies in the $t$-channel are always well below any possible decay thresholds. For the u-channel $\Sigma$ exchange, as a spin-1/2 baryon, is given by
\begin{align}
& S_{\Sigma}(p) = \frac{i(\slashed{p} + M_\Sigma)}{p^2 - M_\Sigma^2}.
\end{align}

\subsection{Phenomenological Form Factors}\label{sec:Phenomenological Form Factors}
A phenomenological form factor is attached to each hadronic vertex~\cite{Wei:2022nqp}. For t-channel meson exchanges, the form factor is
\begin{align}
    f_{M}(q_{M}^{2}) = \left(\frac{\Lambda_{M}^{2}-M_{M}^{2}}{\Lambda_{M}^{2}-q_{M}^{2}}\right)^{2},
\label{eq:ff_meson_app}
\end{align}
where $M_M$ and $q_M$ are the mass and four-momentum of the exchanged meson, and $\Lambda_M$ is the cutoff parameter. For s- and u-channel baryon exchanges, we use the form:
\begin{align}
    f_{B}(p_{x}^{2}) = \left(\frac{\Lambda_{B}^{4}}{\Lambda_{B}^{4}+(p_{x}^{2}-M_{B}^{2})^{2}}\right)^{2}, \quad (x=s,u),
\end{align}
where $M_B$ and $p_x$ are the mass and four-momentum of the exchanged baryon, and $\Lambda_B$ is the corresponding cutoff parameter. Note that the dipole-like forms chosen here are standard in the field~\cite{Huang:2012xj,Wang:2017tpe,Wang:2018vlv} as they provide an effective suppression of the amplitudes at high energies, consistent with the composite nature of hadrons. Our main physical conclusions are not sensitive to this specific choice; the more significant parameters are the cutoff values $\Lambda$, which are determined by the fit to the data.

\subsection{Definitions of Polarization Observables}
\label{sec:Definitions of Polarization Observables}
Besides the unpolarized differential cross-section $d\sigma/d\cos\theta$, this work involves single- and double-polarization observables, as listed in Tab.~\ref{Table1}. The single-polarization observables are the photon beam asymmetry $\Sigma$, the target asymmetry $T$, and the recoil polarization $P$. The double-polarization observables are of the beam-recoil type, namely $C_x$ and $C_z$, which describe the polarization transfer from a circularly polarized photon beam to the recoiling hyperon, and $O_x$ and $O_z$, which describe the polarization transfer from a linearly polarized photon beam. The precise mathematical definitions of these observables in terms of helicity amplitudes, adhere to the standard conventions detailed in Refs.~\cite{Dey:2010fb,Sandorfi:2010uv}, as given below.

The polarization observables are defined as asymmetries of the differential cross-section, which we denote by $\sigma \equiv d\sigma/d\Omega$ for brevity. The notation $\sigma(P_\gamma, P_T, P_R)$ represents the cross-section for a given polarization state of the photon beam ($P_\gamma$), target ($P_T$), and recoil baryon ($P_R$). A value of ``0'' indicates an unpolarized state. The unpolarized cross-section is denoted by $\sigma_0$.

\subsubsection{Single-Polarization Observables}
These observables involve the polarization of one of the particles in the reaction~\cite{Dey:2010fb}.
\begin{align}
\Sigma &= \frac{\sigma(\perp, 0, 0) - \sigma(\parallel, 0, 0)}{\sigma(\perp, 0, 0) + \sigma(\parallel, 0, 0)}, \\
T &= \frac{\sigma(0, +y, 0) - \sigma(0, -y, 0)}{\sigma(0, +y, 0) + \sigma(0, -y, 0)}, \\
P &= \frac{\sigma(0, 0, +y) - \sigma(0, 0, -y)}{\sigma(0, 0, +y) + \sigma(0, 0, -y)}.
\end{align}

\subsubsection{Double-Polarization Observables (Beam-Recoil)}
These observables describe the correlation between the polarization of the photon beam and the recoiling baryon~\cite{Dey:2010fb}.
\begin{align}
C_x &= \frac{\sigma(r, 0, +x) - \sigma(r, 0, -x)}{\sigma(r, 0, +x) + \sigma(r, 0, -x)}, \\
C_z &= \frac{\sigma(r, 0, +z) - \sigma(r, 0, -z)}{\sigma(r, 0, +z) + \sigma(r, 0, -z)}, \\
O_x &= \frac{\sigma(+t, 0, +x) - \sigma(+t, 0, -x)}{\sigma(+t, 0, +x) + \sigma(+t, 0, -x)}, \\
O_z &= \frac{\sigma(+t, 0, +z) - \sigma(+t, 0, -z)}{\sigma(+t, 0, +z) + \sigma(+t, 0, -z)}.
\end{align}
In the definitions above, the coordinate system is defined in the center-of-mass frame where the $\hat{z}$-axis is along the incoming photon momentum, and the $\hat{y}$-axis is normal to the reaction plane ($\hat{y} = \vec{k} \times \vec{q} / |\vec{k} \times \vec{q}|$, where $\vec{k}$ and $\vec{q}$ are the photon and meson momenta). The polarization states are specified as: $\sigma(\perp)$ and $\sigma(\parallel)$ for photons linearly polarized perpendicular and parallel to the reaction plane, respectively. The labels $\pm x, \pm y, \pm z$ indicate the direction of the baryon's spin quantization axis. The state ``r'' denotes a right-handed circularly polarized photon beam, while ``+t'' denotes a photon beam linearly polarized at an angle of $+45^\circ$ with respect to the $\hat{x}$-axis.


\section{FITTING SETTINGS}\label{sec:FITTING SETTINGS}
The fit parameters of this theoretical model are adjusted to match the experimental data in a $\chi^2$ minimization using MINUIT~\cite{James:1975dr,iminuit}. Below, we present our selected settings for the experimental data and fit parameters.

\subsection{Data Base}\label{sec:Data base}

\begin{table}[]
\caption{Experimental data used in the fit. The detailed information of the data presented in the table includes the reactions, observables, experimental collaborations, number, references, and additional weights applied in the fit.}
\label{Table1}
\renewcommand{\arraystretch}{1.03}
{
\begin{tabularx}{\columnwidth}
{ m{3.5cm}<{\centering\arraybackslash} | m{3cm}<{\centering\arraybackslash} | m{3cm}<{\centering\arraybackslash}  m{2cm}<{\centering\arraybackslash}  m{2cm}<{\centering\arraybackslash} | m{2cm}<{\centering\arraybackslash} }

\hline\hline 
  \multicolumn{1}{c}{\textbf{Reaction}} & \multicolumn{1}{c}{\textbf{Observable}} & \multicolumn{1}{c}{\textbf{Collaboration}} & \multicolumn{1}{c}{\textbf{Number}} & \multicolumn{1}{c}{\textbf{Ref.}} & \multicolumn{1}{c}{\textbf{Weight}} \\     
\hline

  \multirow{19}{*}{$\gamma p \to K^+ \Sigma^0$} & \multirow{7}{*}{$d\sigma/dcos\theta$} & SAPHIR 2004 & 660 & ~\cite{Glander:2003jw} & \multirow{7}{*}{1} \\
  && LEPS 2006 & 54 & ~\cite{Kohri:2006yx} & \\ 
  && CLAS 2006 & 1010 & ~\cite{CLAS:2005lui} & \\ 
  && CLAS 2010 & 1288 & ~\cite{CLAS:2010aen} & \\
  && Crystal Ball 2014 & 1115 & ~\cite{Schmieden:2014oba} & \\
  && LEPS 2017 & 44 & ~\cite{LEPS:2017pzl} & \\
  && BGOOD 2021 & 22 & ~\cite{Jude:2020byj} & \\
  \cline{2-6}
  & \multirow{3}{*}{$P$} & SAPHIR 2004 & 16 & ~\cite{Glander:2003jw} & \multirow{3}{*}{1} \\
  && GRAAL 2007 & 8 & ~\cite{Lleres:2007tx} & \\
  && CLAS 2010 & 280 & ~\cite{CLAS:2010aen} & \\
  \cline{2-6}
  & \multirow{4}{*}{$\Sigma$} & LEPS 2006 & 30 & ~\cite{Kohri:2006yx} & \multirow{4}{*}{1} \\
  && GRAAL 2007 & 42 & ~\cite{Lleres:2007tx} & \\
  && CLAS 2016 & 127 & ~\cite{CLAS:2016wrl} & \\
  && LEPS 2017 & 12 & ~\cite{LEPS:2017pzl} & \\
  \cline{2-6}
  & \multirow{1}{*}{$T$} & CLAS 2016 & 127 & ~\cite{CLAS:2016wrl}  & \multirow{1}{*}{2} \\
  \cline{2-6}
  & \multirow{1}{*}{$C_x$} & CLAS 2007 & 70 & ~\cite{CLAS:2006pde} & \multirow{1}{*}{3} \\
  \cline{2-6}
  & \multirow{1}{*}{$C_z$} & CLAS 2007 & 63 & ~\cite{CLAS:2006pde} & \multirow{1}{*}{3} \\
  \cline{2-6}
  & \multirow{1}{*}{$O_x$} & CLAS 2016 & 127 & ~\cite{CLAS:2016wrl} & \multirow{1}{*}{3} \\
  \cline{2-6}
  & \multirow{1}{*}{$O_z$} & CLAS 2016 & 127 & ~\cite{CLAS:2016wrl} & \multirow{1}{*}{2} \\
  
  \hline
  \multirow{13}{*}{$\gamma p \to K^0 \Sigma^+$} & \multirow{4}{*}{$d\sigma/dcos\theta$} & SAPHIR 2005 & 120 & ~\cite{Lawall:2005np} & \multirow{4}{*}{7} \\
  && CBELSA 2008 & 72 & ~\cite{CBELSATAPS:2007oqn} & \\ 
  && A2 2013 & 50 & ~\cite{A2:2013cqk} \\ 
  && CBELSA 2012 & 72 & ~\cite{CBELSATAPS:2011gly} & \\ 
  \cline{2-6}
  & \multirow{5}{*}{$P$} & SAPHIR 2005 & 10 & ~\cite{Lawall:2005np} & \multirow{5}{*}{2} \\
  && CBELSA 2008 & 23 & ~\cite{CBELSATAPS:2007oqn} \\ 
  && CLAS 2013 & 78 & ~\cite{CLAS:2013owj} \\ 
  && A2 2013 & 32 & ~\cite{A2:2013cqk} \\ 
  && CLAS 2024 & 21 & ~\cite{CLAS:2024bzi} \\ 
  \cline{2-6}
  & \multirow{1}{*}{$\Sigma$} & CLAS 2024 & 21 & ~\cite{CLAS:2024bzi} & \multirow{1}{*}{5} \\
  \cline{2-6}
  & \multirow{1}{*}{$T$} & CLAS 2024 & 21 & ~\cite{CLAS:2024bzi} & \multirow{1}{*}{5} \\
  \cline{2-6}
  & \multirow{1}{*}{$O_x$} & CLAS 2024 & 21 & ~\cite{CLAS:2024bzi} & \multirow{1}{*}{5} \\
  \cline{2-6}
  & \multirow{1}{*}{$O_z$} & CLAS 2024 & 21 & ~\cite{CLAS:2024bzi} & \multirow{1}{*}{5} \\

  \hline
  \multicolumn{1}{r}{\textbf{In total}}& & & \textbf{5784} & \multicolumn{1}{r}{\textbf{}} & \\
  
\hline\hline
\end{tabularx} 
}
\end{table} 

The experimental data used in our fit are listed in Tab.~\ref{Table1}. We have compiled nearly all available experimental data for the $\gamma p \to K^+ \Sigma^0$ and $\gamma p \to K^0 \Sigma^+$ reactions. However, we exclude certain datasets, including the differential cross-section data for $\gamma p \to K^0 \Sigma^+$ from the A2 2019~\cite{A2:2018doh}, the photon beam asymmetry data for $\gamma p \to K^0 \Sigma^+$ from the CBELSA 2014~\cite{Schmieden:2014oba}, and some older data, due to issues such as inconsistencies with others, sparsity or larger errors. To concentrate on the regions with potential contributions from exchanges of the molecules, we select the experimental data within the center-of-mass energy range from the $K \Sigma$ threshold up to 2400 MeV. The new polarization observables for $\gamma p \to K^0 \Sigma^+$ from CLAS 2024~\cite{CLAS:2024bzi} have also been included.

To address the significant variation in the number of available data points across different observables, we employ a weighting procedure in the $\chi^2$ minimization~\cite{Ronchen:2022hqk,Clymton:2021wof}. This is a standard and often necessary practice in complex coupled-channel analyses to prevent observables with thousands of data points (e.g., differential cross-sections) from completely dominating the fit, thereby ensuring that sparser but physically constraining polarization data are also adequately described. The weight for each observable, listed in Tab.~\ref{Table1}, is multiplied into its corresponding term $\chi^2_{observable}$ to construct the total $\chi^2_{weight}$, which is employed in the fitting process. The final values for these weights are not uniquely determined by a rigid formula; rather, they were chosen through an iterative process of adjustment, with the goal of achieving a reasonably balanced and high-quality description across all included observables. Notably, the overall fitted results and main physical conclusions appear to be reasonably stable against moderate variations of these weights.

\subsection{Fit Parameters}\label{sec:Fit parameters}
Here, we introduce the fit parameters of our theoretical model, which are listed in Tabs.~\ref{Table3} and ~\ref{Table4}. First, the electromagnetic and hadronic coupling constants of the $K_1(1270)$, molecules, general $N^*$ and $\Delta^*$ resonances, are treated as free parameters that need to be fitted. Since the reaction amplitudes are only sensitive to the products of electromagnetic and hadronic coupling constants, we make the products as the fit parameters instead of individual coupling constants, which are shown in Tab.~\ref{Table3}. It is necessary to note that for the specific calculations of the reactions $\gamma p \to K^+ \Sigma^0$ and $\gamma p \to K^0 \Sigma^+$, the products $g_{\gamma N R} g_{K \Sigma R}$ should be multiplied by the corresponding isospin factor $\tau$.

Second, to reduce the number of fit parameters, we implement the following settings for the masses $M_R$, widths $\Gamma_R$, and phases $\phi_R$ of molecules and general resonances. For molecules, the widths and phases are treated as fit parameters, while the masses are fixed, as mentioned in Sec.~\ref{sec:FORMALISM}. For four-star general resonances, only the masses of $\Delta(1910)1/2^+$ and the widths of $N(1720)3/2^+$, $N(1900)3/2^+$, $\Delta(1910)1/2^+$ are treated as fit parameters due to their large ranges recorded in RPP~\cite{ParticleDataGroup:2024cfk} and relatively significant effects on fitted results. The masses and widths of other four-star general resonances are fixed according to RPP~\cite{ParticleDataGroup:2024cfk}. For three-star and two-star general resonances, all masses and widths are treated as fit parameters.

Lastly, the cutoff parameters $\Lambda$ in the phenomenological form factors attached in each hadronic vertex are also treated as fit parameters. We merge some of them to reduce the number of fit parameters. Specifically, we use the same cutoff parameter $\Lambda_t$ for the $t$-channel $K$ and $K^*(892)$ exchanges, and the same cutoff parameter $\Lambda_s$ for the $s$-channel ground states $N$ and $\Delta$ exchanges. Additionally, we merge the cutoff parameters of the molecules and general resonances located below the $K \Sigma$ threshold. For the molecules and general resonances above the threshold, the cutoff parameters are divided into ten groups based on their category and spin-parity, which are shown in Tab.~\ref{Table4}.

In summary, our theoretical model contains a total of 77 fit parameters, which is a relatively reduced number, that need to be adjusted to match the experimental data through the fitting program.

\section{RESULTS AND DISCUSSION}\label{sec:RESULTS AND DISCUSSIONS}

\subsection{Fitted Results}\label{sec:Fitted results}

We construct $\chi^{2}_{weight}$ with weights shown in Tab.~\ref{Table1}, then determine the fitted values of the model's free parameters by minimizing $\chi^{2}_{weight}$ with MINUIT. Owing to the large number of fitting parameters and sparse experimental data for some observables, the fitting process produces several convergent results. We selected the most representative set as our final results. The results of all 77 fit parameters are listed in Tabs.~\ref{Table3} and ~\ref{Table4}, and the corresponding values of $\chi^2$ are listed in Tab.~\ref{Table2}.

\begin{table}[h]
\caption{Specific values of molecular and resonant parameters. The fitted values of free parameters are presented with uncertainties, while the values of the other parameters are fixed. The values in the brackets below general resonances' masses and widths are corresponding values advocated by RPP~\cite{ParticleDataGroup:2024cfk}.}
\label{Table3}
\renewcommand{\arraystretch}{0.96}
{
\begin{tabularx}{\columnwidth}
{ >{\centering\arraybackslash}X  >{\centering\arraybackslash}X  >{\centering\arraybackslash}X >{\centering\arraybackslash}X  >{\centering\arraybackslash}X  >{\centering\arraybackslash}X }

\hline\hline 
  \multicolumn{1}{c}{\textbf{Molecule}} & \multicolumn{1}{c}{\textbf{$M_R [MeV]$}} & \multicolumn{1}{c}{\textbf{$\Gamma_R [MeV]$}} & \multicolumn{1}{c}{\textbf{$g_{\gamma N R}^{(1)}g_{K \Sigma R}$}} & \multicolumn{1}{c}{\textbf{$g_{\gamma N R}^{(2)}g_{K \Sigma R}$}} & \multicolumn{1}{c}{\textbf{$\phi _R$}} \\     
\hline

{$N(1535)1/2^-$} & 1535 & $450 $ & $-0.157 \pm 0.004$ & & $ -0.190 \pm 0.021$\\
$N(1875)3/2^-$ & 1875 & $450 $ & $-10.900 \pm 0.190$ & $12.230 \pm 0.220$ & $ -2.473 \pm 0.017 $ \\
$N(2080)1/2^-$ & 2080 & $203 \pm 9$ & $-0.050 \pm 0.003$ & & $2.756 \pm 0.031$ \\
$N(2080)3/2^-$ & 2080 & $144 \pm 7$ & $0.630 \pm 0.040$ & $-0.880 \pm 0.040$ & $0.069 \pm 0.034$ \\
$N(2270)1/2^-$ & 2270 & $261 \pm 10$ & $-0.040 \pm 0.003$ & & $3.685 \pm 0.032$ \\
$N(2270)3/2^-$ & 2270 & $450 $ & $2.013 \pm 0.034$ & $-2.320 \pm 0.040$ & $-11.203 \pm 0.026$ \\
$N(2270)5/2^-$ & 2270 & $450 $ & $-0.574 \pm 0.013$ & $-0.610 \pm 0.040$ & $0.998 \pm 0.021$ \\

\hline
\multicolumn{1}{c}{\textbf{Resonance}} & \multicolumn{1}{c}{\textbf{$M_R [MeV]$}} & \multicolumn{1}{c}{\textbf{$\Gamma_R [MeV]$}} & \multicolumn{1}{c}{\textbf{$g_{\gamma N R}^{(1)}g_{K \Sigma R}$}} & \multicolumn{1}{c}{\textbf{$g_{\gamma N R}^{(2)}g_{K \Sigma R}$}} & \multicolumn{1}{c}{\textbf{}} \\     
\hline
$N(1675) \ 5/2^-$ & 1675 & $145$ & $0.492 \pm 0.035$ & $2.630 \pm 0.120$ & \\
\addlinespace[-8pt]
{\scriptsize****} & {\scriptsize$[1665\sim1680]$} & {\scriptsize$[130\sim160]$} & & & \\

$N(1710) \ 1/2^+$ & 1710 & $140$ & $0.193 \pm 0.015$ & & \\
\addlinespace[-8pt]
{\scriptsize****} & {\scriptsize$[1680\sim1740]$} & {\scriptsize$[80\sim200]$} & & & \\

$N(1720) \ 3/2^+$ & 1720 & $414 \pm 13$ & $0.859 \pm 0.033$ & $-0.480 \pm 0.050$ & \\
\addlinespace[-8pt]
{\scriptsize****} & {\scriptsize$[1680\sim1750]$} & {\scriptsize$[150\sim400]$} & & & \\

$N(1880) \ 1/2^+$ & $1858 \pm 7$ & $404 \pm 16$ & $0.567 \pm 0.026$ & & \\
\addlinespace[-8pt]
{\scriptsize***} & {\scriptsize$[1830\sim1930]$} & {\scriptsize$[200\sim400]$} & & & \\

$N(1900) \ 3/2^+$ & 1920 & $155 \pm 3$ & $0.189 \pm 0.005$ & $-0.360 \pm 0.014$ & \\
\addlinespace[-8pt]
{\scriptsize****} & {\scriptsize$[1890\sim1950]$} & {\scriptsize$[100\sim320]$} & & & \\

$\Delta (1600) \ 3/2^+$ & 1570 & $250$ & $-1.804\pm0.020$ & $2.719\pm0.040$ & \\
\addlinespace[-8pt]
{\scriptsize****} & {\scriptsize$[1500\sim1640]$} & {\scriptsize$[200\sim300]$} & & & \\

$\Delta (1700) \ 3/2^-$ & 1710 & $300$ & $-0.727\pm0.144$ & $-0.012\pm0.162$ & \\
\addlinespace[-8pt]
{\scriptsize****} & {\scriptsize$[1690\sim1730]$} & {\scriptsize$[220\sim380]$} & & & \\

$\Delta (1900) \ 1/2^-$ & $1853\pm2$ & $161 \pm 8$ & $0.053\pm0.003$ & & \\
\addlinespace[-8pt]
{\scriptsize***} & {\scriptsize$[1840\sim1920]$} & {\scriptsize$[180\sim320]$} & & & \\

$\Delta (1910) \ 1/2^+$ & $1950\pm1$ & $400$ & $-0.953\pm0.009$ & & \\
\addlinespace[-8pt]
{\scriptsize****} & {\scriptsize$[1850\sim1950]$} & {\scriptsize$[200\sim400]$} & & & \\

$\Delta (1920) \ 3/2^+$ & $1913\pm2$ & $178 \pm 8$ & $0.111\pm0.006$ & $0.040\pm0.023$ & \\
\addlinespace[-8pt]
{\scriptsize***} & {\scriptsize$[1870\sim1970]$} & {\scriptsize$[240\sim360]$} & & & \\

$\Delta (1930) \ 5/2^-$ & $1937\pm2$ & $286 \pm 14$ & $-1.080\pm0.069$ & $0.647\pm0.144$ & \\
\addlinespace[-8pt]
{\scriptsize***} & {\scriptsize$[1900\sim2000]$} & {\scriptsize$[200\sim400]$} & & & \\

$\Delta (1940) \ 3/2^-$ & $1940\pm1$ & $500$ & $-7.280\pm0.092$ & $9.671\pm0.115$ & \\
\addlinespace[-8pt]
{\scriptsize**} & {\scriptsize$[1940\sim2060]$} & {\scriptsize$[300\sim500]$} & & & \\

\hline
\multicolumn{1}{c}{\textbf{}} & \multicolumn{1}{c}{\textbf{$M_{K_1} [MeV]$}} & \multicolumn{1}{c}{\textbf{$\Gamma_{K_1} [MeV]$}} & \multicolumn{1}{c}{\textbf{$g_{\gamma K^+ K_1^+}g_{\Sigma^0 p K_1^+}^{(1)}$}} & \multicolumn{1}{c}{\textbf{$g_{\gamma K^+ K_1^+}g_{\Sigma^0 p K_1^+}^{(2)}$}} & \multicolumn{1}{c}{\textbf{$g_{\gamma K^0 K_1^0}g_{\Sigma^+ p K_1^0}^{(1)}$}} \\     
\hline

$K_1(1270) \ 1^+$ & 1253 & $90$ & $1.060\pm0.150$ & $-1.876\pm0.166$ &  $-0.082\pm0.022$ \\

\hline\hline
\end{tabularx}
}
\end{table} 

\begin{table}[]
\caption{Fitted values of cutoff parameters (in MeV). The exchanged particles listed below share the same cutoff value.}
\label{Table4}
\renewcommand{\arraystretch}{0.9}
{
\begin{tabularx}{\columnwidth}
{ >{\centering\arraybackslash}X  >{\centering\arraybackslash}X  >{\centering\arraybackslash}X >{\centering\arraybackslash}X  >{\centering\arraybackslash}X }

\hline\hline 
{\large$\Lambda_t$} & {\large$\Lambda_{K_1}$} & {\large$\Lambda_u$} & {\large$\Lambda_s$} & {\large$\Lambda_1$} \\
\addlinespace[-7pt]
{\tiny$K,K^*(892)$} & {\tiny$K_1(1270)$} & {\tiny$\Sigma$} & {\tiny$N,\Delta$} & {\tiny$N(2080) 1/2^-,N(2270) 1/2^-$} \\
\hline
$667\pm1$ & $767\pm21$ & $700\pm1$ & $985\pm3$ & $2200\pm60$ \\
\hline\hline \\
\hline\hline

{\large$\Lambda_2$} & {\large$\Lambda_3$} & {\large$\Lambda_4$} & {\large$\Lambda_5$} & {\large$\Lambda_6$} \\   
\addlinespace[-7pt]
{\tiny$N(1710),N(1880)$} & {\tiny$N(1720),N(1900)$} & {\tiny$N(1875) 3/2^-, N(2080) 3/2^-$} & {\tiny$N(2270) 5/2^-$} & {\tiny$\Delta(1900)$} \\
\addlinespace[-8pt]
&  & {\tiny$N(2270) 3/2^-$} &  &  \\
\hline
$2000\pm3$ & $1395\pm24$ & $837\pm6$ & $1150\pm4$ & $1750\pm70$ \\
\hline\hline \\
\hline\hline

 {\large$\Lambda_7$} & {\large$\Lambda_8$} & {\large$\Lambda_9$} & {\large$\Lambda_{10}$} & {\large$\Lambda_{11}$} \\   
\addlinespace[-7pt]
{\tiny$\Delta(1910)$} & {\tiny$\Delta(1920)$} & {\tiny$\Delta(1700), \Delta(1940)$} & {\tiny$\Delta(1930)$} & {\tiny$N(1535) 1/2^-, N(1675), \Delta (1600) $} \\
\hline
$2000\pm7$ & $873\pm15$ & $1281\pm6$ & $750\pm1$ & $1700\pm1$ \\
 \hline\hline
 
\end{tabularx}
}
\end{table} 

Tabs.~\ref{Table3} and ~\ref{Table4} present the specific values of 77 fit parameters and some other fixed parameters in our theoretical model. It should be noted that that during the fitting process, we observed that the widths of $N(1535)1/2^-$, $N(1875)3/2^-$, $N(2270)3/2^- \&\ 5/2^-$, $\Delta(1910)1/2^+$ and $\Delta(1940)3/2^-$ tend to be larger; however, they have no significant impact on the fitted results. Consequently, the widths of $N(1535)1/2^-$, $N(1875)3/2^-$, $N(2270)3/2^- \&\ 5/2^-$ are set at 450 MeV, while the widths of $\Delta(1910)1/2^+$ and $\Delta(1940)3/2^-$ are set at the upper limits of the width ranges recorded in RPP~\cite{ParticleDataGroup:2024cfk}. Apart from these parameters, the other fit parameters have convergent fitted values with associated errors.

For the molecules, they do not necessarily correspond to the established particles listed in RPP~\cite{ParticleDataGroup:2024cfk}. Consequently, their parameters are not strictly constrained by RPP~\cite{ParticleDataGroup:2024cfk}. As calculated in Refs.~\cite{Lin:2018kcc,Ben:2024qeg}, the widths of these molecules exhibit a significant dependence on the cutoff parameters, and the range of widths for the molecules can cover the possible width range of the general nucleon excited states. Thus, it is difficult for us to provide accurate predictions for both the total width and the coupling constants of the molecules. In this context, setting the width of $N(1535)1/2^-$, $N(1875)3/2^-$, $N(2270)3/2^- \&\ 5/2^-$ to 450 MeV serves as a reference value that reflects the fit's preference for a large width, given that the fitted results are not sensitive to this specific number. In particular, the large width of the sub-threshold $N(1535)1/2^-$ may effectively incorporate the influence of other sub-threshold resonances. Additionally, the fitted values of molecules listed in Tab.~\ref{Table3} are deemed consistent with the calculations in Refs.~\cite{Lin:2018kcc,Ben:2024qeg} and regarded as reference. 

For the general resonances in the $s$-channel, Tab.~\ref{Table3} lists the fitted values of their fit parameters along with their basic information. Besides the coupling constants, some widths and masses are treated as fit parameters, with fitted values basically falling within the range recorded in RPP~\cite{ParticleDataGroup:2024cfk}. For $K_1 (1270)$, only three products of coupling constants serve as independent free parameters, and their fitted values are also presented in Tab.~\ref{Table3}. 

Tab.~\ref{Table4} presents the fitted values of the free cutoff parameters, with the exchanged particles listed below sharing the same cutoff value, as noted in Sec.~\ref{sec:Fit parameters}.

We have also gotten the correlation matrix for the 77 fit parameters. As is often the case in a high-dimensional parameter space, most parameters exhibit no significant relationship with each other, apart from certain typical correlations. For instance, a strong negative correlation is often observed between the two coupling constants of a single resonance, such as for the $N(1875)3/2^-$. This is a natural consequence of the partial cancellation or substitution between the two terms in controlling the resonance's total contribution. Negative correlations can also appear between coupling constants of different resonances, such as between the $N(2080)1/2^-$ and $N(2270)1/2^-$ states, suggesting their contributions can be partially substitutive in certain kinematic regions. However, given the large number of parameters, it is challenging to draw further significant physical conclusions from the correlations alone, and thus a detailed analysis is not provided here. 

\begin{table}[]
\caption{The $\chi^2$ values for the reactions $\gamma p \to K^+ \Sigma^0$ and $K^0 \Sigma^+$. The $\chi^2/N_{data}$ for each individual observable, as well as the $\chi^2/(N_{data}-N_{par.})$ for the total data, are presented, where $N_{data}$ and $N_{par.}$ denote the number of experimental data points and fit parameters, respectively.}
\label{Table2}
\renewcommand{\arraystretch}{1.0}
{
\begin{tabularx}{\columnwidth}
{ >{\centering\arraybackslash}X  >{\centering\arraybackslash}X  >{\centering\arraybackslash}X }

\hline\hline
\textbf{Reaction} & \textbf{Observable \ ($N_{data}$)} & \textbf{$\chi^2/N_{data}$} \\
\hline
\multirow{9}{*}{\large$\gamma p \to K^+ \Sigma^0$} & $d\sigma / dcos\theta$ \ (4193) & 1.055 \\
   & $P$ \ (304) & 1.852 \\
   & $\Sigma$ \ (211) & 5.376 \\
   & $T$ \ (127) & 1.748 \\
   & $C_x$ \ (70) & 1.881 \\
   & $C_z$ \ (63) & 1.801 \\
   & $O_x$ \ (127) & 2.501 \\
   & $O_z$ \ (127) & 1.754 \\
   & \textbf{In total} \ (5222) & \textbf{1.365}  \\

\hline
\multirow{7}{*}{\large$\gamma p \to K^0 \Sigma^+$} & $d\sigma / dcos\theta$ \ (314) & 1.519 \\
   & $P$ \ (164) & 1.861 \\
   & $\Sigma$ \ (21) & 1.360 \\
   & $T$ \ (21) & 1.684 \\
   & $O_x$ \ (21) & 2.110 \\
   & $O_z$ \ (21) & 0.787 \\
   & \textbf{In total} \ (562) & \textbf{1.614} \\

\hline
\addlinespace[-5pt]
\multirow{2}{*}{\textbf{\large In total} \ (5784)} & & \multirow{2}{*}{\large $\chi^2/(N_{data}-N_{par.}) = $ \textbf{1.408}} \\
 \addlinespace[-5pt]
 & & \\
 \hline\hline
\end{tabularx}
}
\end{table} 

Tab.~\ref{Table2} clearly illustrates the fit quality of the final fitted results. Here, $\chi^2$ presents an unweighted chi-squared statistic, while $N_{data}$ and $N_{par.}$ indicate the number of experimental data points and free parameters, respectively. Nearly all the $\chi^2/N_{data}$ values of the observables in the two reactions are around or below 2, indicating a high fitting quality for each observable. The only exception is the $\Sigma$ of the reaction $\gamma p \to K^+ \Sigma^0$, where the $\chi^2/N_{data}$ is relatively high. However, this is primarily due to the rather small error bars in the data points, while the fitting quality remains good, as shown in Fig.~\ref{Σp}. Furthermore, the total $\chi^2/(N_{data}-N_{par.})$ is 1.408, demonstrating the overall high quality of the fitted results. In conclusion, both in terms of individual observables and the overall picture, the quality of our fitted results is satisfactory. Achieving this is challenging for coupled-channel fits that involve two different reactions and data sets from various measurements, demonstrating the effectiveness of our theoretical model.

\begin{table}[]
\caption{The convergent fitted values for the masses of the molecules above the $K\Sigma$ threshold. These error-inclusive values come from the further fitting with the fitted results presented above as initial values.}
\label{Table5}
\renewcommand{\arraystretch}{1.0}
{
\begin{tabularx}{\columnwidth}
{ >{\centering\arraybackslash}X  >{\centering\arraybackslash}X  >{\centering\arraybackslash}X >{\centering\arraybackslash}X  >{\centering\arraybackslash}X >{\centering\arraybackslash}X}

\hline\hline 
{$M_{N(1875)3/2^-}$} & {$M_{N(2080)1/2^-}$} & {$M_{N(2080)3/2^-}$} & {$M_{N(2270)1/2^-}$} & {$M_{N(2270)3/2^-}$}& {$M_{N(2270)5/2^-}$} \\
\hline
$1896\pm4$ & $2047\pm5$ & $2005\pm5$ & $2408\pm5$ & $2258\pm3$ & $2216\pm4$ \\
\hline\hline
 
\end{tabularx}
}
\end{table} 

Moreover, as mentioned in Sec.~\ref{sec:FORMALISM}, the masses of the molecules are fixed in our theoretical model. To verify the stability of the fitted results with fixed molecular masses presented above, we further use this set of fitted values as initial values to perform the fitting with the molecular masses above the $K\Sigma$ threshold released. The convergent fitted values for the masses of $N(1875)3/2^-$, $N(2080)1/2^- \&\ 3/2^-$, $N(2270)1/2^- , 3/2^- \&\ 5/2^-$ are listed in Tab.~\ref{Table5}. The variation range for most of these masses is within 100 MeV. Notably, for $N(2270)3/2^-$, its mass varies by only 12 MeV, indicating a particularly strong tendency for $N(2270)3/2^-$ to contribute in this region. In contrast, the variation in the mass of $N(2270)1/2^-$ is relatively larger, and as shown in the following Fig.~\ref{plot1}, its contribution is also comparatively small, suggesting that the experimental data does not strongly favor it.

\subsection{Cross-Sections}\label{sec:cross-sections} 

\begin{figure}[]
    \includegraphics[scale=0.65]{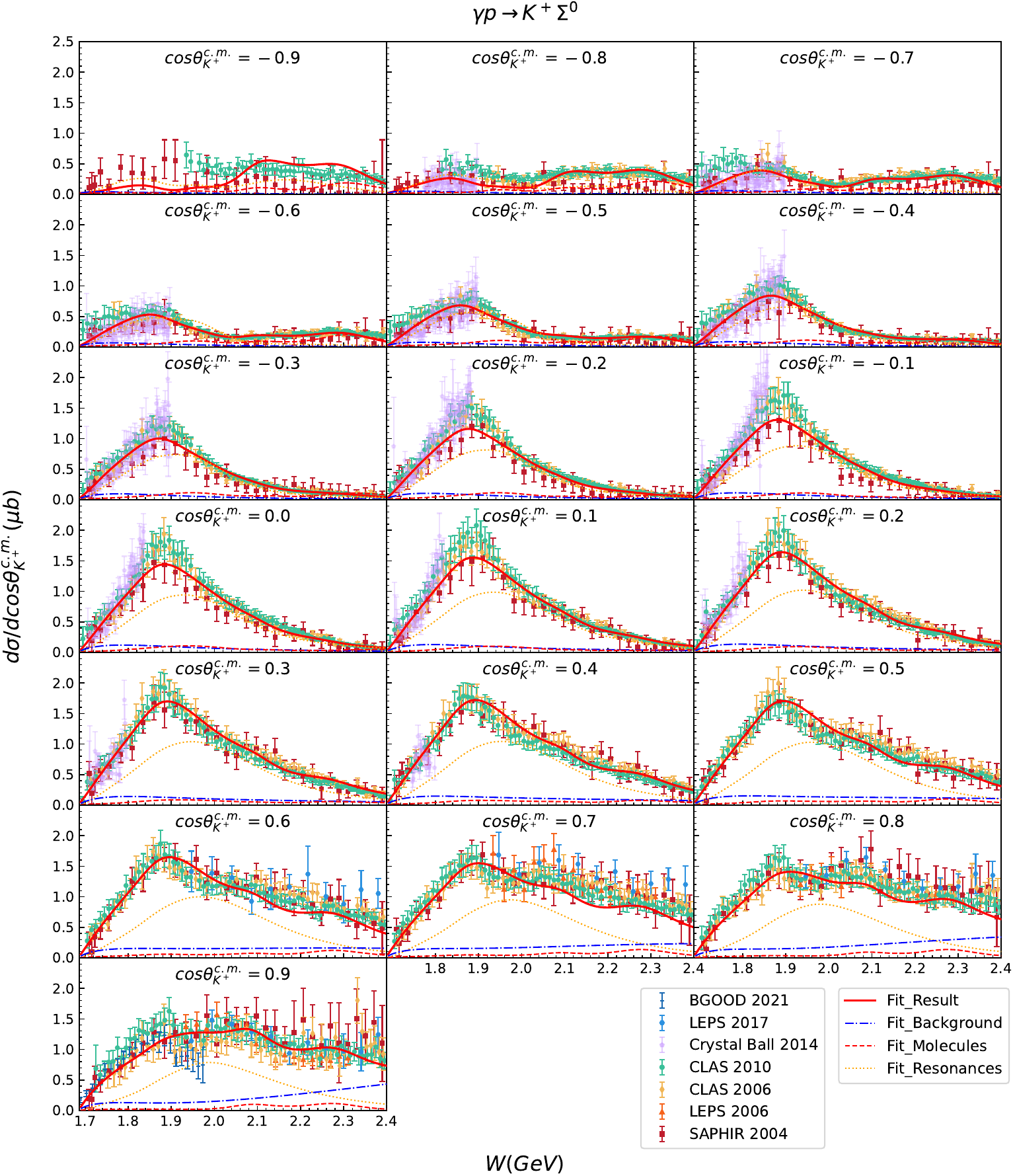}
    \caption{Differential cross-sections for $\gamma p \to K^+ \Sigma^0$ as a function of the total center-of-mass energy $W$. The collaborations for the experimental data are listed in the legend, with detailed information provided in Tab.~\ref{Table1}. The red solid line denotes our theoretical result based on the parameters in Tabs.~\ref{Table3} and ~\ref{Table4}, while the other three dashed lines represent contributions from $s$-channel molecule exchanges, $s$-channel general resonance exchanges, and the background (all other terms).}
    \label{dsigmap}
\end{figure}

\begin{figure}[]
    \includegraphics[scale=0.65]{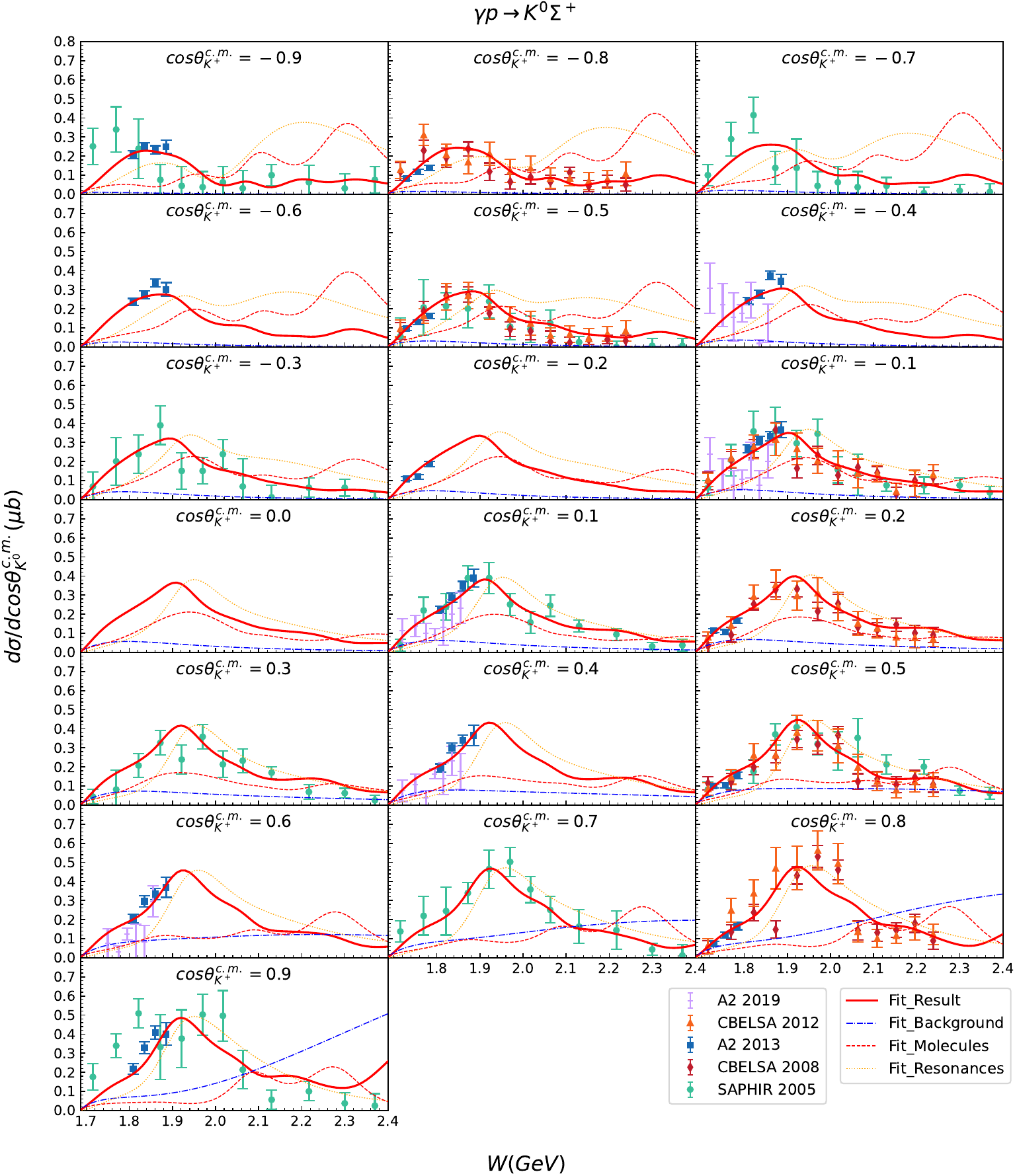}
    \caption{Differential cross-sections for $\gamma p \to K^0 \Sigma^+$ as a function of the total center-of-mass energy $W$. Except for A2 2019~\cite{A2:2018doh}, which was excluded from the fitting database due to inconsistencies, all other experimental data are available in Tab.~\ref{Table1}. The notation for theoretical results follows that in Fig.~\ref{dsigmap}.}
    \label{dsigma0}
\end{figure}

Figs.~\ref{dsigmap} and ~\ref{dsigma0} present the theoretical and experimental results for the differential cross-sections of $\gamma p \to K^+ \Sigma^0$ and $\gamma p \to K^0 \Sigma^+$, respectively. Our theoretical numerical results, corresponding to the parameters listed in Tab.~\ref{Table3}, are compared with nearly all available experimental data. Additionally, the individual contributions from $s$-channel molecule exchanges, $s$-channel general resonance exchanges, and all the other terms—collectively referred to as the background—are displayed to facilitate the analysis of the reaction mechanisms.

\begin{figure}[]
\centering
\subfigure[The total cross-section for $\gamma p \to K^+ \Sigma^0$. Except for SAPHIR 1998~\cite{SAPHIR:1998fev} and ABBHHM 1969~\cite{Aachen-Berlin-Bonn-Hamburg-Heidelberg-Muenchen:1969pjo}, other collaborations of experimental data can be found in Tab.~\ref{Table1}. The blue thick dashed line represents calculated result of the HFF-P3 model in Ref.~\cite{Clymton:2021wof}. Notation for our theoretical numerical results is as in Fig.~\ref{dsigmap}. Note that all data of the total cross-section shown in this figure were not used in the fitting database.]{
\includegraphics[width=0.75\textwidth]{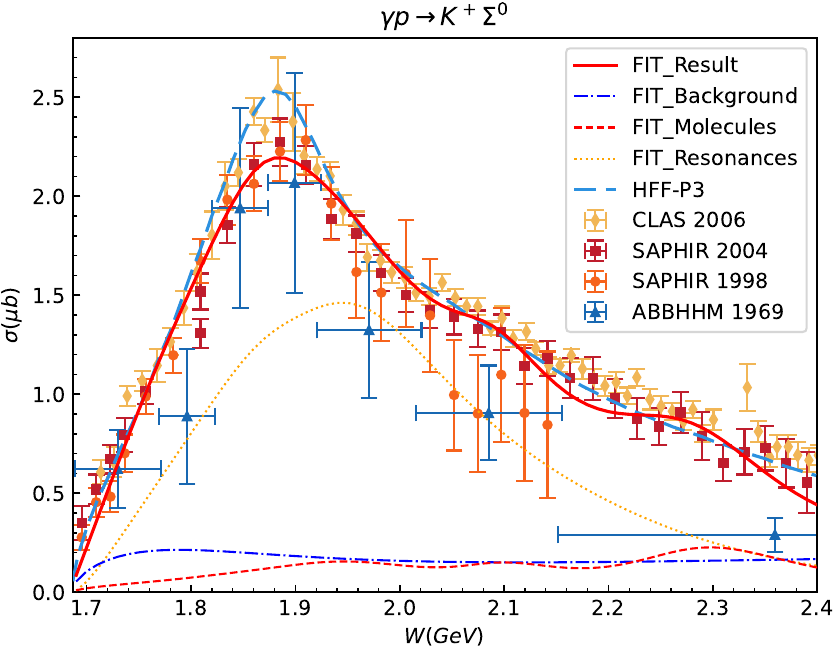}
\label{σtp}
} 

\subfigure[Molecules]{
\includegraphics[width=0.48\textwidth]{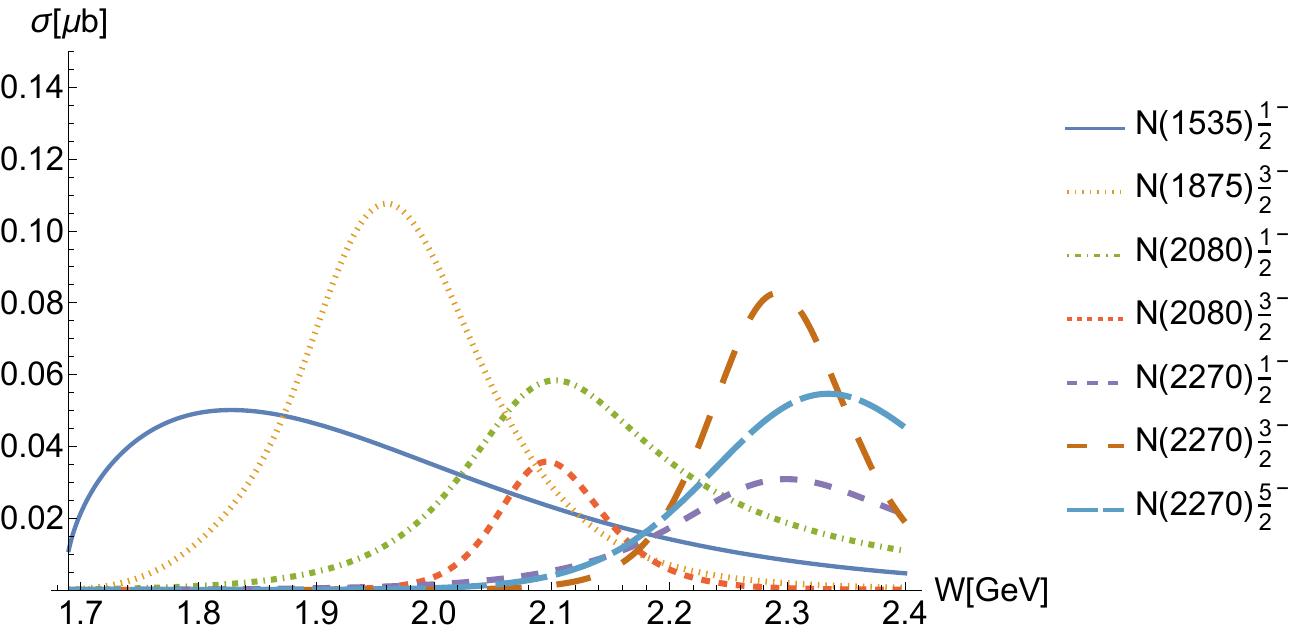}
\label{plot1}
} 
\subfigure[General $N^*$ resonances]{
\includegraphics[width=0.48\textwidth]{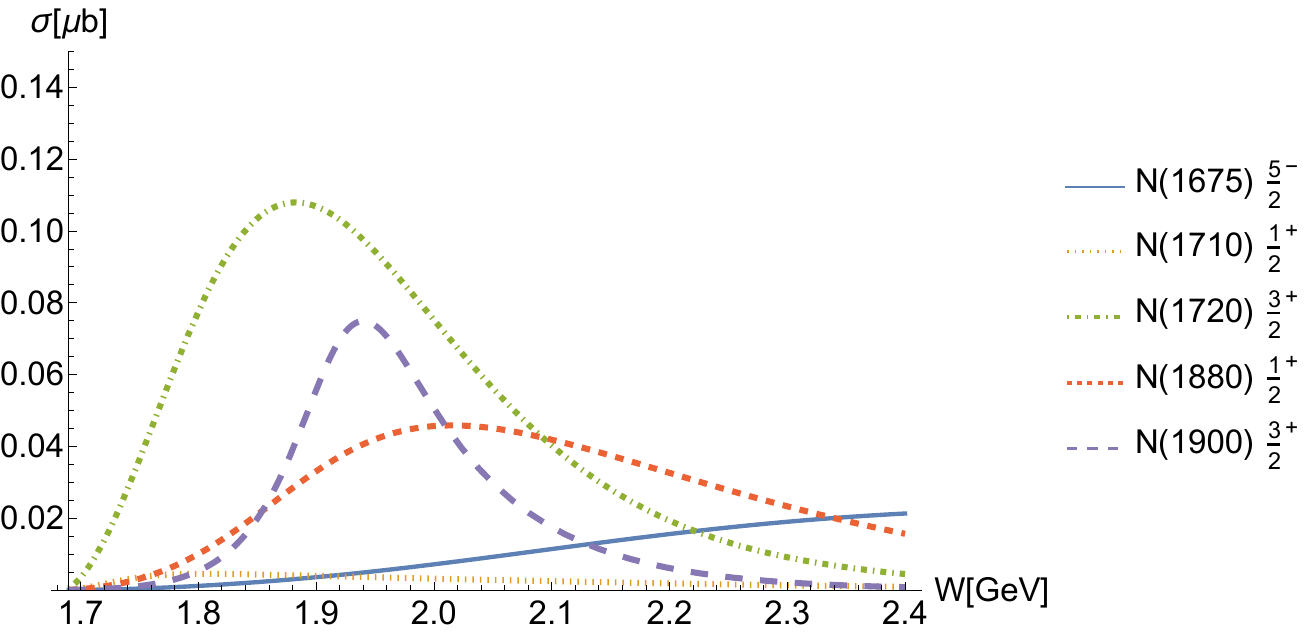}
\label{plot2}
} 

\subfigure[$\Delta^*$ resonances]{
\includegraphics[width=0.48\textwidth]{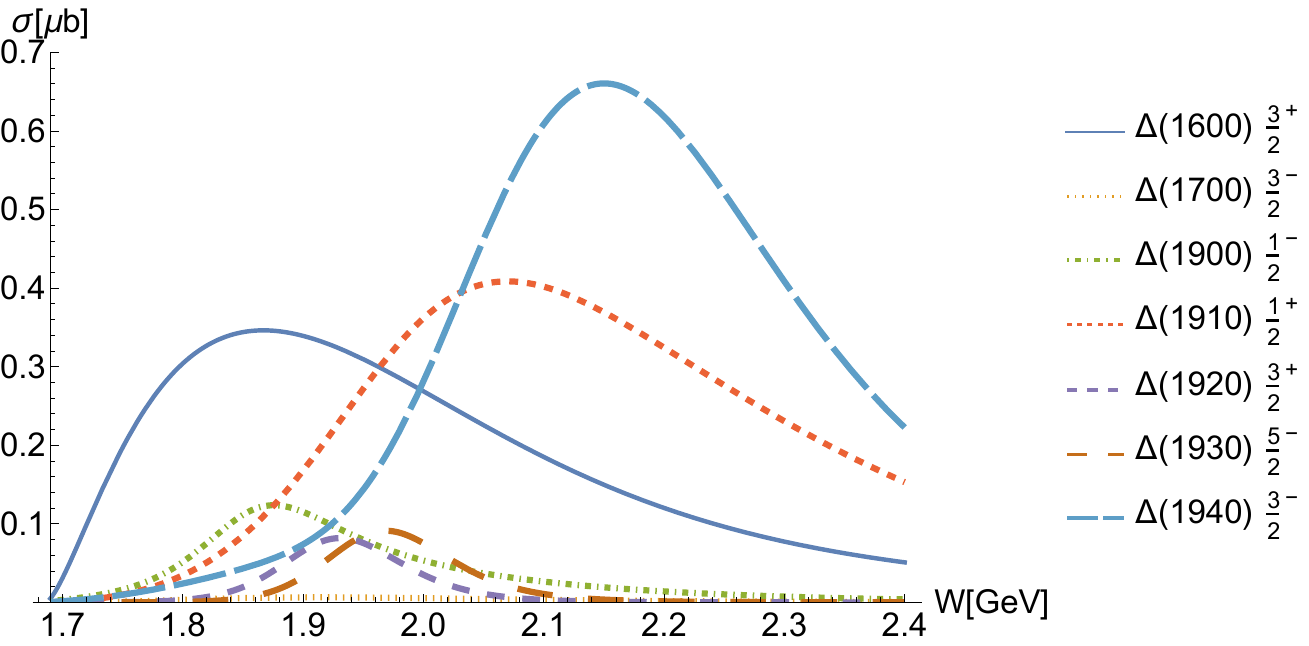}
\label{plot3}
} 
\subfigure[Background]{
\includegraphics[width=0.48\textwidth]{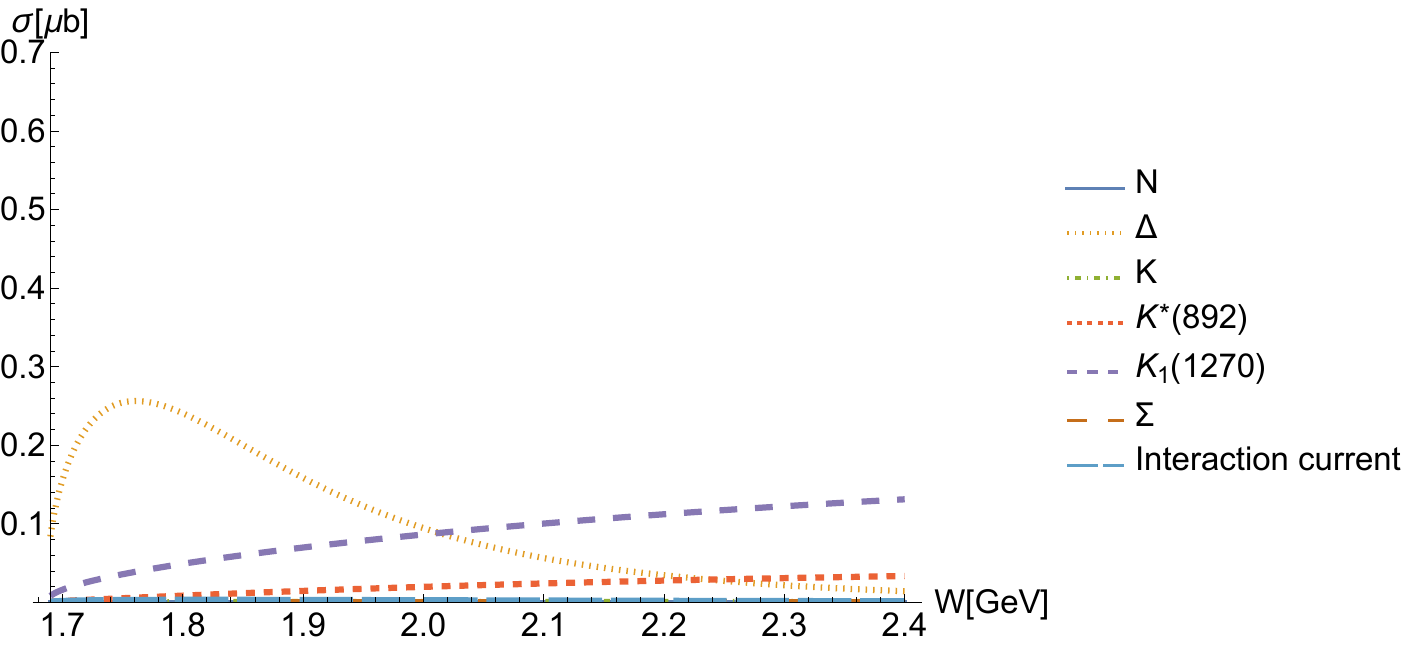}
\label{plot4}
} 

\caption{The total cross-section for $\gamma p \to K^+ \Sigma^0$, along with the individual contributions from single particle exchanges labeled on the right.}
\label{sigmap}
\end{figure}

\begin{figure}[]
\centering
\subfigure[The total cross-section for $\gamma p \to K^0 \Sigma^+$. Except for CLAS 2005~\cite{Klein:2005mv}, other collaborations of experimental data can be found in Tab.~\ref{Table1}. Notation for numerical results of our theory and the HFF-P3 model is as in Fig.~\ref{sigmap}. For $K^0 \Sigma^+$, the upper limit of the HFF-P3 model's result presented in Ref.~\cite{Clymton:2021wof} is 2150 MeV(marked with a black dashed line). Note that all data of the total cross-section shown in this figure were not used in the fitting database.]{
\includegraphics[width=0.75\textwidth]{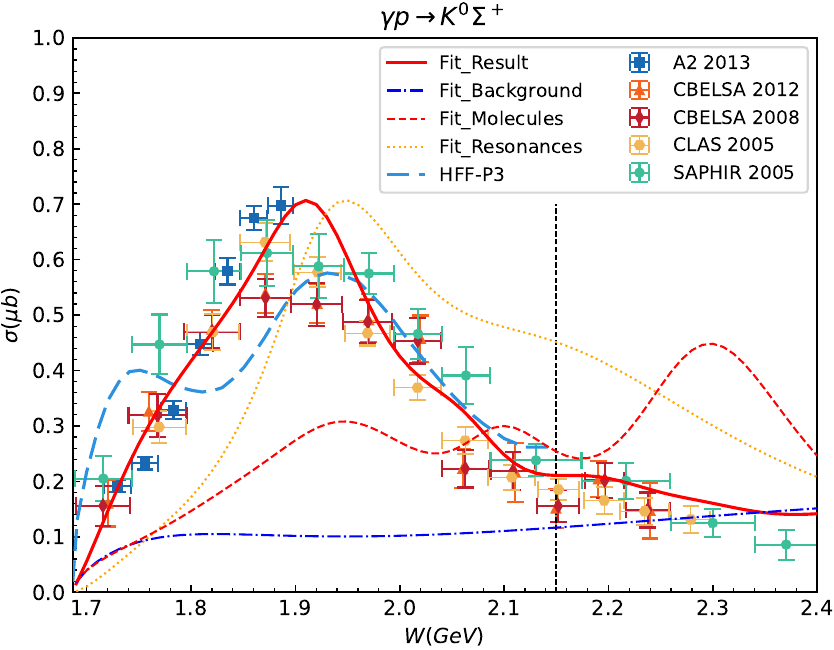}
\label{σ0}
} 

\subfigure[Molecules]{
\includegraphics[width=0.48\textwidth]{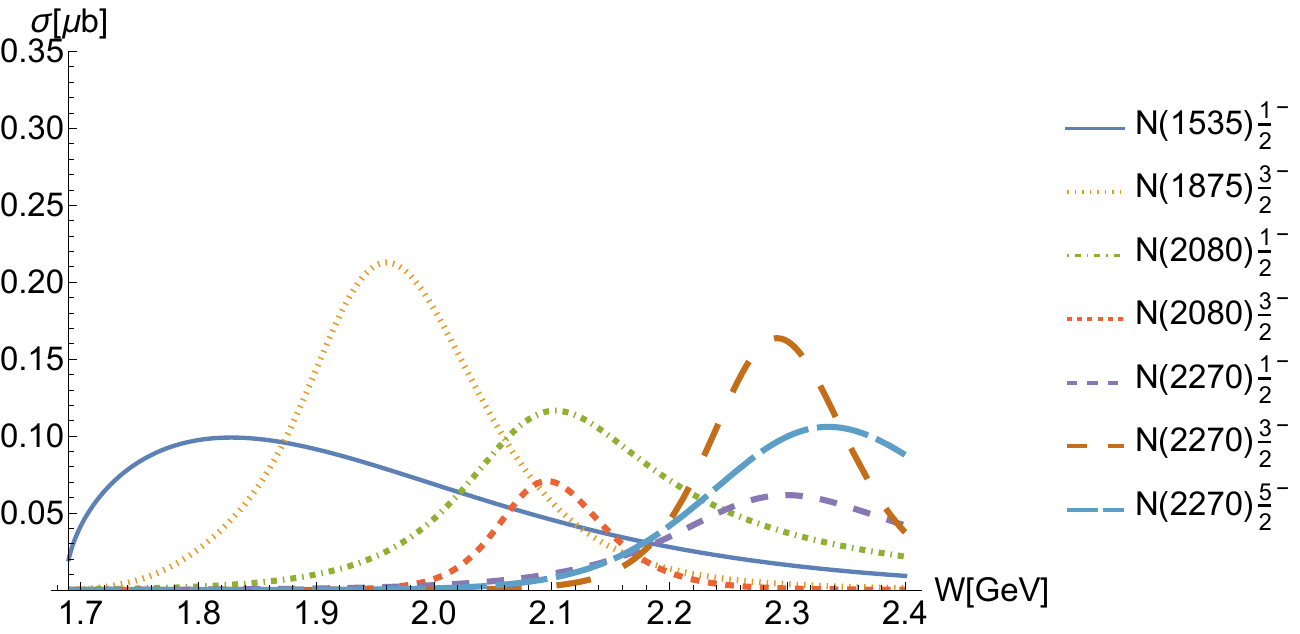}
\label{plot5}
} 
\subfigure[General $N^*$ resonances]{
\includegraphics[width=0.48\textwidth]{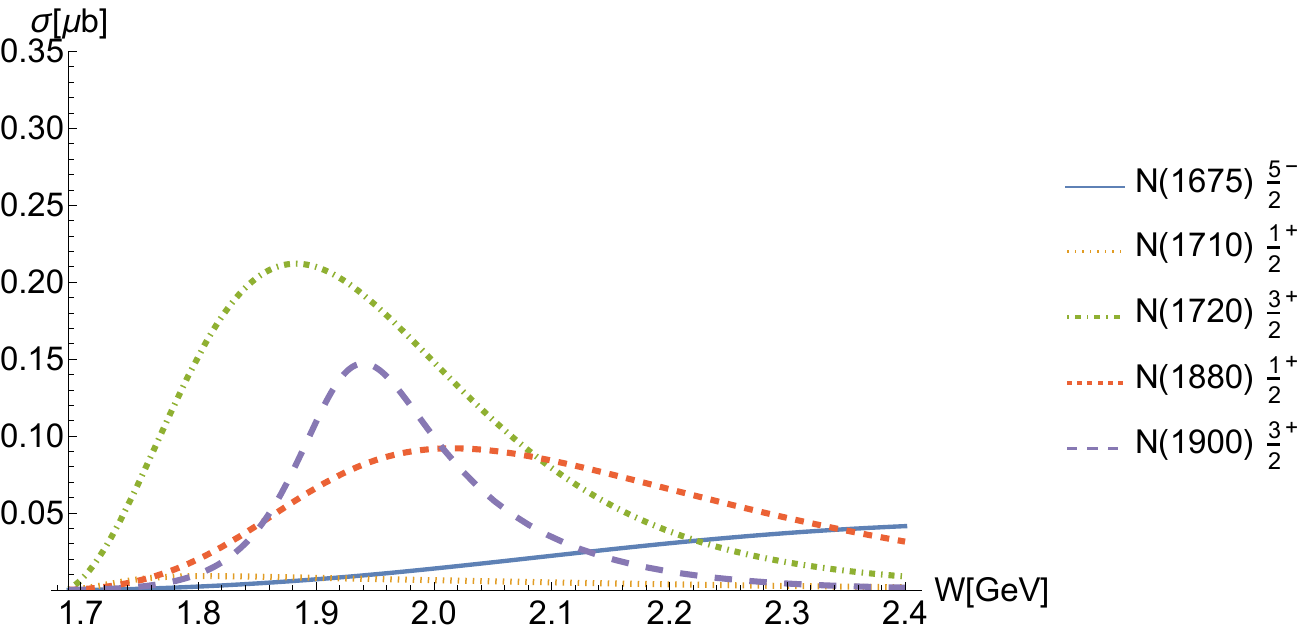}
\label{plot6}
} 

\subfigure[$\Delta^*$ resonances]{
\includegraphics[width=0.48\textwidth]{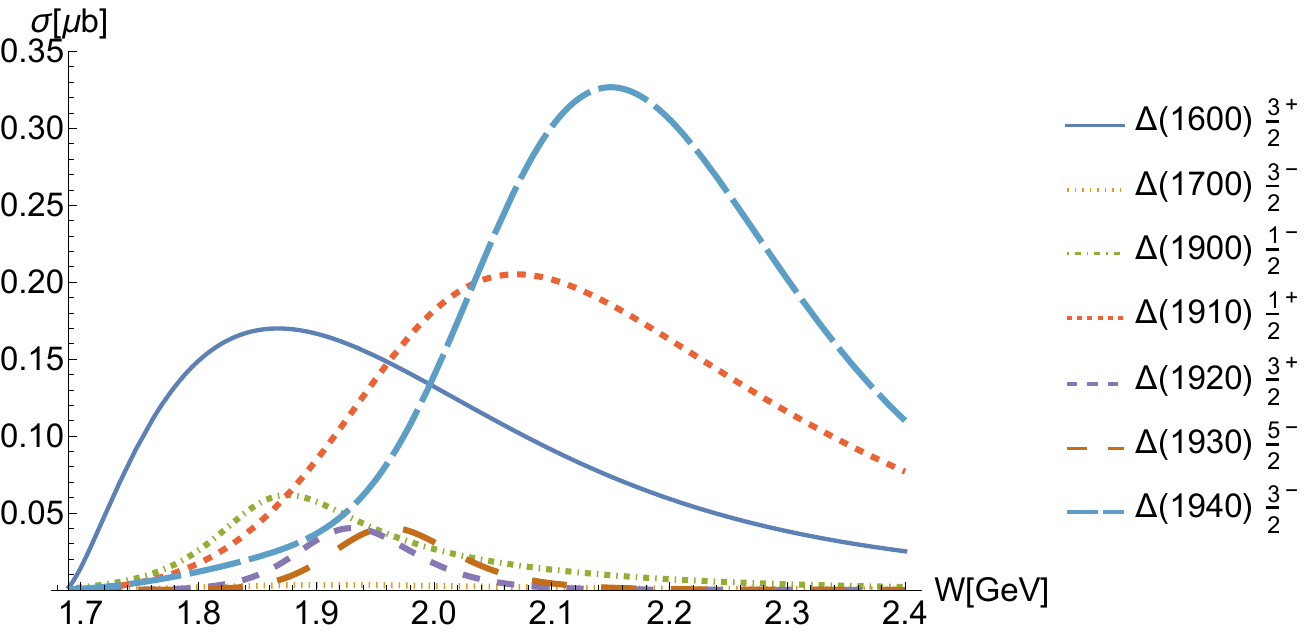}
\label{plot7}
} 
\subfigure[Background]{
\includegraphics[width=0.48\textwidth]{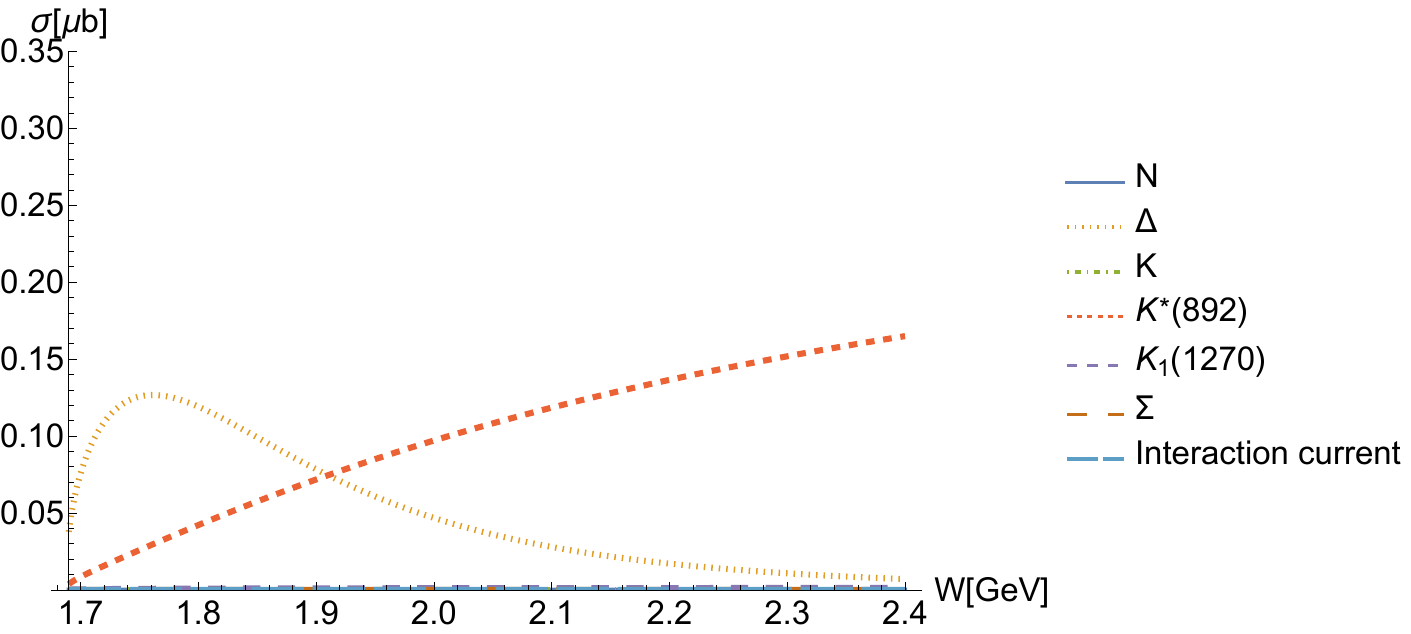}
\label{plot8}
} 

\caption{The total cross-section for $\gamma p \to K^0 \Sigma^+$, along with the individual contributions from single particle exchanges labeled on the right.}
\label{sigma0}
\end{figure}

Figs.~\ref{sigmap} and ~\ref{sigma0} present the theoretical and experimental results for the total cross-sections of these two reactions, along with the individual contributions from single particle exchanges displayed below. The experimental data for the total cross-sections shown in the figures were not used in the fitting database, and are just compared with our theoretical predicted results. In addition, the calculated results of the HFF-P3 model in Ref.~\cite{Clymton:2021wof} are included for further comparison. The work in Ref.~\cite{Clymton:2021wof} provides a comprehensive analysis of nearly all available data for the four possible isospin channels of $K \Sigma$ photoproduction using a covariant isobar model. It is highly representative; according to Ref.~\cite{Clymton:2021wof}, among the three models, ``the model HFF-P3 shows the best agreement with the experimental data (lowest $\chi^2$) from all but the $\gamma n \to K^0 \Sigma^0$ channel.'' Therefore, the calculated results of the HFF-P3 model are particularly valuable for comparison with our results.

Overall, the experimental data for both the differential and total cross-sections of the two reactions are well described. As shown in Tab.~\ref{Table2}, the $\chi^2/N_{data}$ values of the differential cross-sections are 1.055 for $\gamma p \to K^+ \Sigma^0$ and 1.519 for $\gamma p \to K^0 \Sigma^+$, indicating excellent agreement. Considering the differences in the amount of experimental data for the two reactions, this is a challenging yet satisfactory outcome, demonstrating the effectiveness of our theoretical model and the settings of weights. Additionally, from Figs.~\ref{dsigmap} to ~\ref{sigma0}, we can also see that contributions from the $s$-channel molecule exchanges are essential, indicating that the effects of the molecules are potentially significant in the $\gamma p \to K \Sigma$ reactions. Moreover, we will discuss several other important features of the results below.

For the reaction $\gamma p \to K^+ \Sigma^0$, contributions from the $s$-channel $\Delta^*$ resonance exchanges are dominant, as illustrated in detail in Fig.~\ref{plot3}. In fact, this is a reasonable expectation, which we will analyze in detail later by comparing the cross-sections and isospin factors of the two reactions. Regarding the background, as shown in Fig.~\ref{plot4}, $s$-channel proton exchange, $t$-channel $K$ exchange, $u$-channel $\Sigma$ exchange and the interaction current have little contributions. The ground state $\Delta$ exchange has a relatively significant contribution, similar to other $\Delta^*$ resonances. And the $t$-channel $K^* (892)$ and $K_1 (1270)$ exchanges provide considerable contributions of differential cross-sections at the forward angles in the high energy regions, as shown in Fig.~\ref{dsigmap}.

As for the $s$-channel molecule exchanges, as shown in Fig.~\ref{plot1}, the $N(1875)3/2^-$ exchange provides the largest contributions among molecules. Alongside $N(1535)1/2^-$, exchanges of these two molecules contribute across a wide energy range due to their relatively large widths. Together with contributions from $s$-channel general resonance exchanges, they help construct the overall structure of the cross-sections, particularly the peak at $W \approx 1900$ MeV. In addition, $N(2080)1/2^- \&\ 3/2^-$ and $N(2270)1/2^- , 3/2^- \&\ 5/2^-$ exchanges are mainly responsible for the peak structures around $W$ = 2080 and 2270 MeV, respectively, observable at both the backward and forward angular regions of the differential cross-sections in Fig.~\ref{dsigmap}, as well as in the total cross-section shown in Fig.~\ref{sigmap}. The contributions from these molecules with different spins are roughly comparable, as illustrated in Fig.~\ref{plot1}, showing no obvious preference for any particular spin. 

In Fig.~\ref{sigmap}, we compare the total cross-section result from our theoretical model (red thick solid line) with that from the HFF-P3 model (blue thick dashed line) in Ref.~\cite{Clymton:2021wof}. Our result exhibits distinct peaks around $W$ = 2080 and 2270 MeV, while the HFF-P3 result appears smoother. This discrepancy indicates the significant effects of molecules within our model. In Sec.~\ref{sec:INTRODUCTION}, we have mentioned that the bump structures near $W$ = 1875, 2080 and 2270 MeV in differential cross-sections for $\gamma p \to K^+ \Sigma^0$, serve as one of the motivations for investigating the effects of the molecules in $\gamma p \to K^+ \Sigma^0 / K^0 \Sigma^+$ reactions. The final fitted results indicate that these peak structures do contain significant contributions from the molecules. 

From $\gamma p \to K^+ \Sigma^0$ to $\gamma p \to K^0 \Sigma^+$, the isospin factor $\tau$ of $g_{K \Sigma \Delta}$ changes from $\sqrt{2}$ to $1$, while the $\tau$ of $g_{K \Sigma N}$ changes from $-1$ to $\sqrt{2}$. This is inclined to suggest that contributions from the $s$-channel $\Delta^*$ resonance exchanges are more substantial for the reaction $\gamma p \to K^+ \Sigma^0$, based on a simple comparison of the magnitudes of cross-sections for the two reactions shown in Figs.~\ref{sigmap} and ~\ref{sigma0}, respectively. Meanwhile, the contributions from $N^*$ and $\Delta^*$ resonance exchanges have become comparable for the reaction $\gamma p \to K^0 \Sigma^+$, as depicted in Figs.~\ref{plot5} to ~\ref{plot7}, due to the variation of the isospin factor $\tau$. Therefore, if we want to investigate $N^*$ resonances, the $K^0 \Sigma^+$ reaction seems to be more important due to the amplified effects on isospin factors.

For the reaction $\gamma p \to K^0 \Sigma^+$, in terms of background, as shown in Fig.~\ref{plot8}, the contribution from $K_1(1270)$ exchange becomes negligible, while the contribution from $K^*(892)$ exchange is much larger than its in the $\gamma p \to K^+ \Sigma^0$ channel as shown in Fig.~\ref{plot4}. Furthermore, the effects of the molecules are more pronounced. Fig.~\ref{dsigma0} illustrates the substantial interference effects between the contributions from $s$-channel general resonance exchanges and molecule exchanges. The interference effects arise not only from isospin factors but also from the phase factors considered in the calculation of molecules, and these are one of the important reasons for the significant differences in the magnitudes of the cross-sections for $\gamma p \to K^+ \Sigma^0$ and $\gamma p \to K^0 \Sigma^+$. Aside from these, most of the contribution characteristics of cross-sections for $\gamma p \to K^0 \Sigma^+$ are similar to those for $\gamma p \to K^+ \Sigma^0$. In Fig.~\ref{sigma0}, we also compare the total cross-section result from our theoretical model with that from the HFF-P3 model, which is provided up to 2150 MeV in Ref.~\cite{Clymton:2021wof}. And our result exhibits the additional variability around $W$ = 2080 MeV, due to the effects of the molecules.

However, it is clear that the experimental data for the reaction $\gamma p \to K^0 \Sigma^+$ are much sparser compared to these for the reaction $\gamma p \to K^+ \Sigma^0$. We improved the fitted results for $\gamma p \to K^0 \Sigma^+$ by adjusting the weights during the fitting procedure, but we hope to obtain more experimental data for $\gamma p \to K^0 \Sigma^+$ in the future to strengthen the constraints on the theoretical models.

\subsection{Polarization Observables}\label{sec:Polarization observables}

\begin{figure}[]
    \includegraphics[scale=0.44]{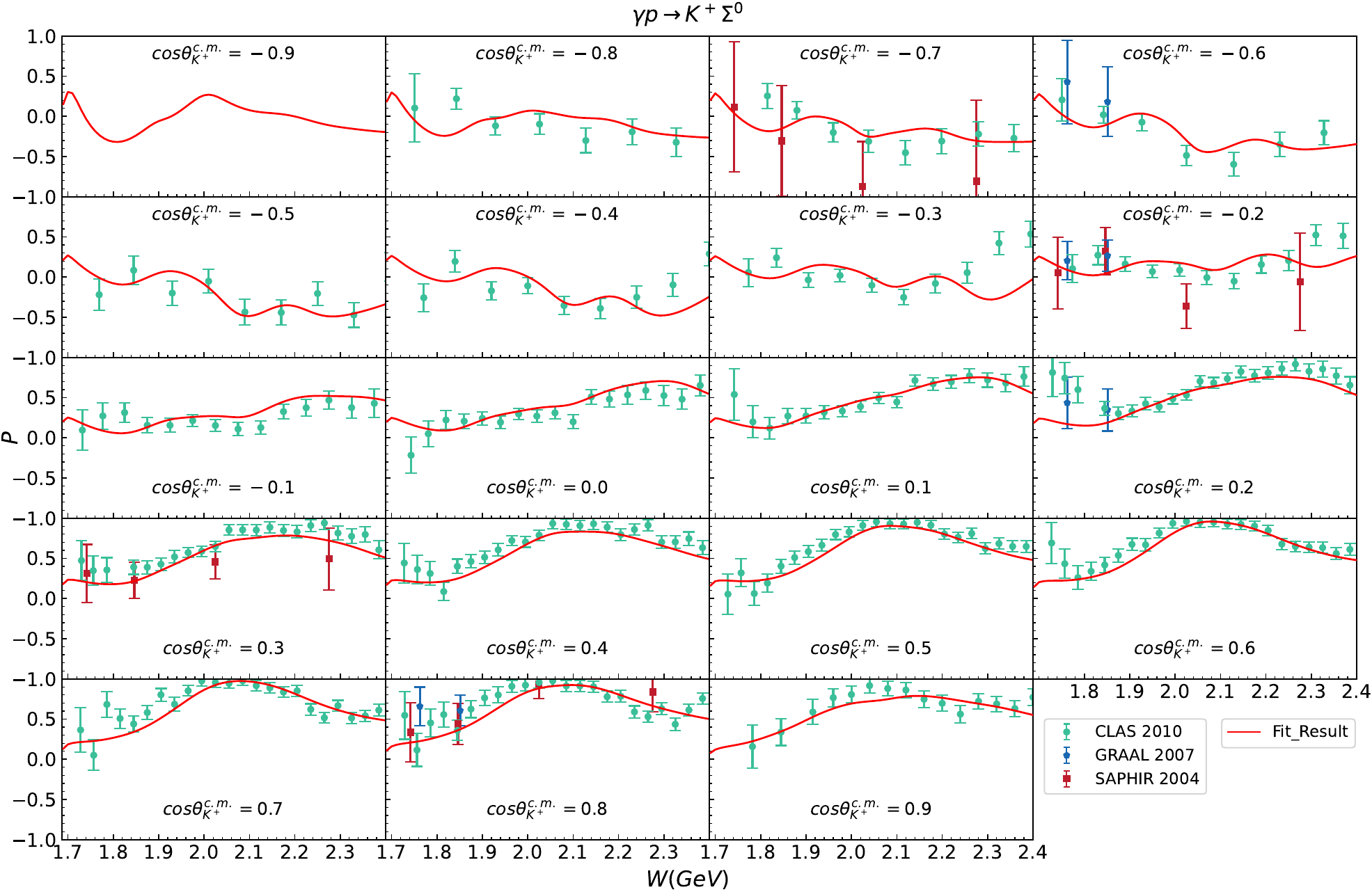}
    \caption{Recoil polarization $P$ for $\gamma p \to K^+ \Sigma^0$ obtained in our theoretical calculations corresponding to the parameters listed in Tabs.~\ref{Table3} and ~\ref{Table4}, compared with the experimental data shown in Tab.~\ref{Table1}.}
    \label{Pp}
\end{figure}

\begin{figure}[]
    \includegraphics[scale=0.44]{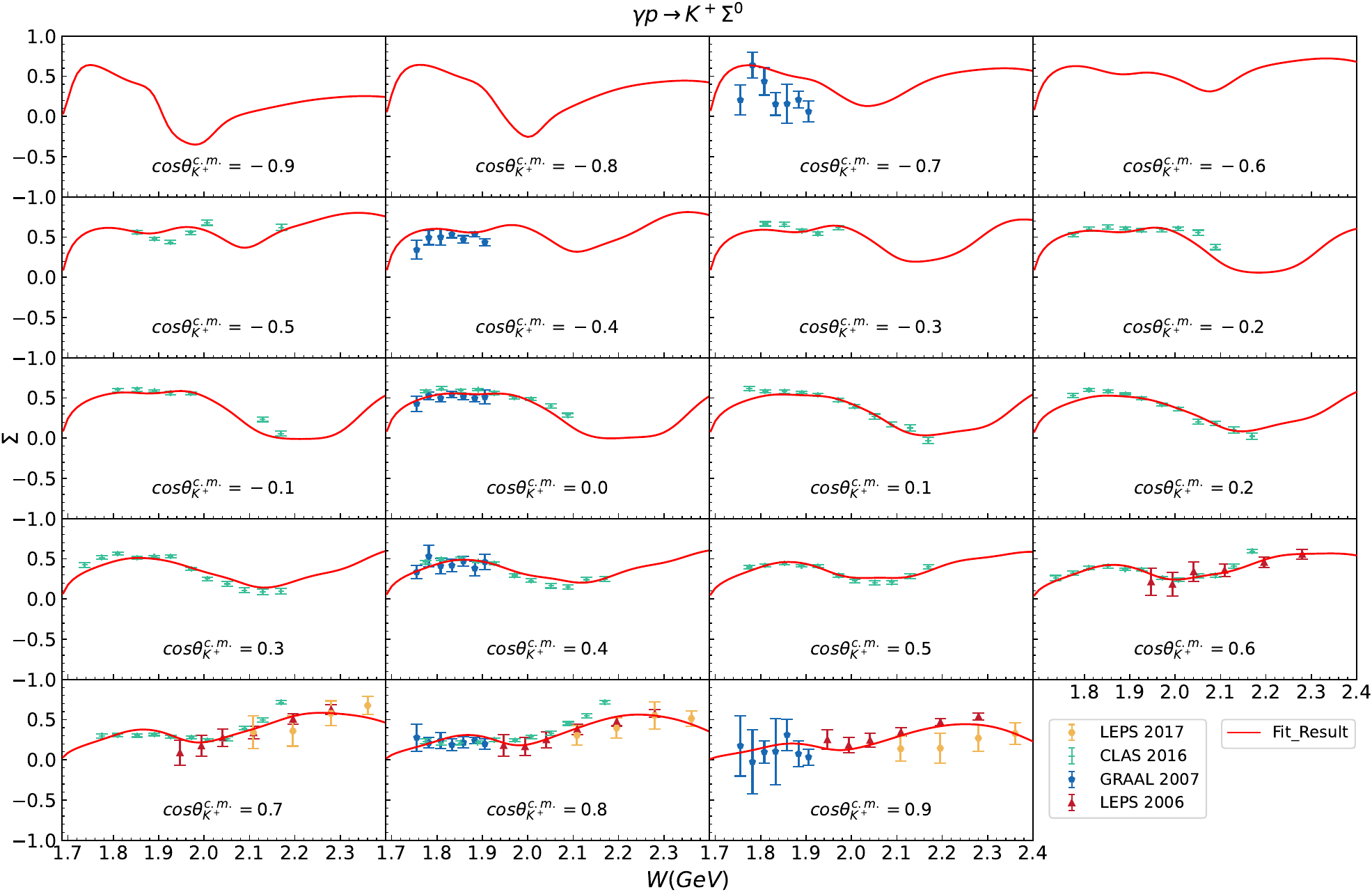}
    \caption{Photon beam asymmetry $\Sigma$ for $\gamma p \to K^+ \Sigma^0$. Notation for the theoretical and experimental results is shown in the legend.}
    \label{Σp}
\end{figure}

\begin{figure}[]
    \includegraphics[scale=0.44]{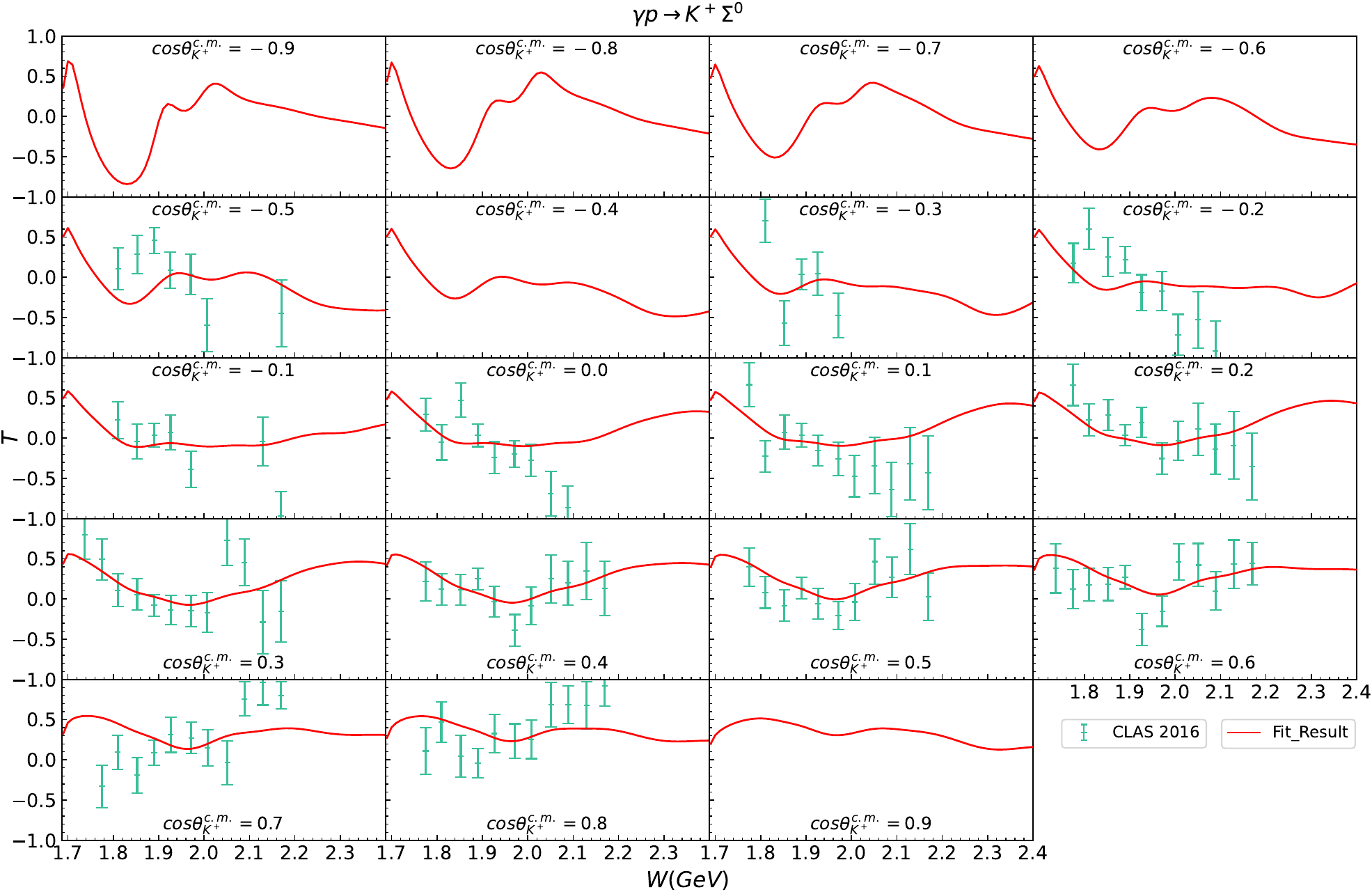}
    \caption{Target asymmetry $T$ for $\gamma p \to K^+ \Sigma^0$. Notation for the theoretical and experimental results is shown in the legend.}
    \label{Tp}
\end{figure}

\begin{figure}[]
    \includegraphics[scale=0.44]{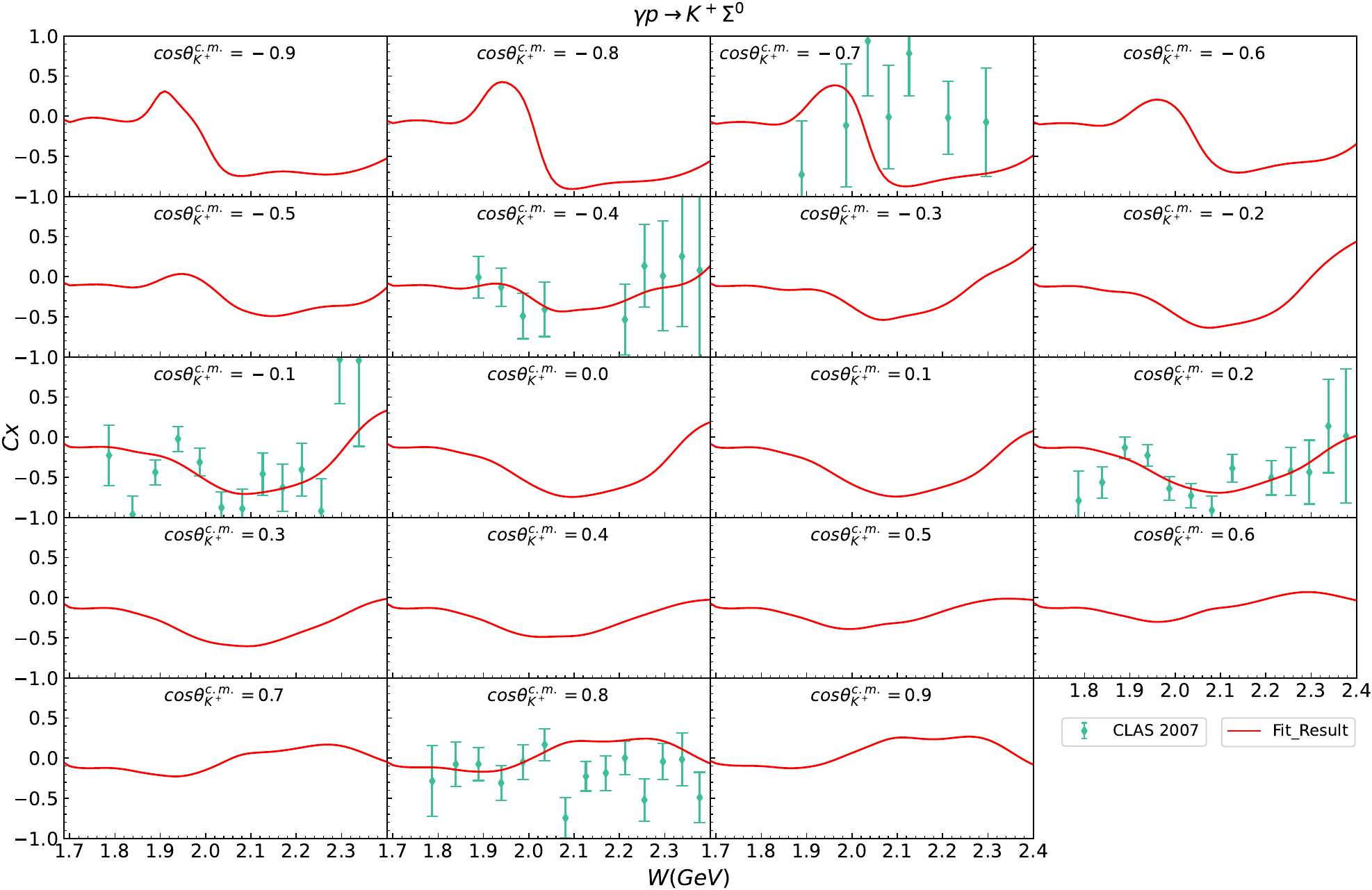}
    \caption{Beam-recoil double polarization $C_x$ for $\gamma p \to K^+ \Sigma^0$. Notation for the theoretical and experimental results is shown in the legend.}
    \label{Cxp}
\end{figure}

\begin{figure}[]
    \includegraphics[scale=0.44]{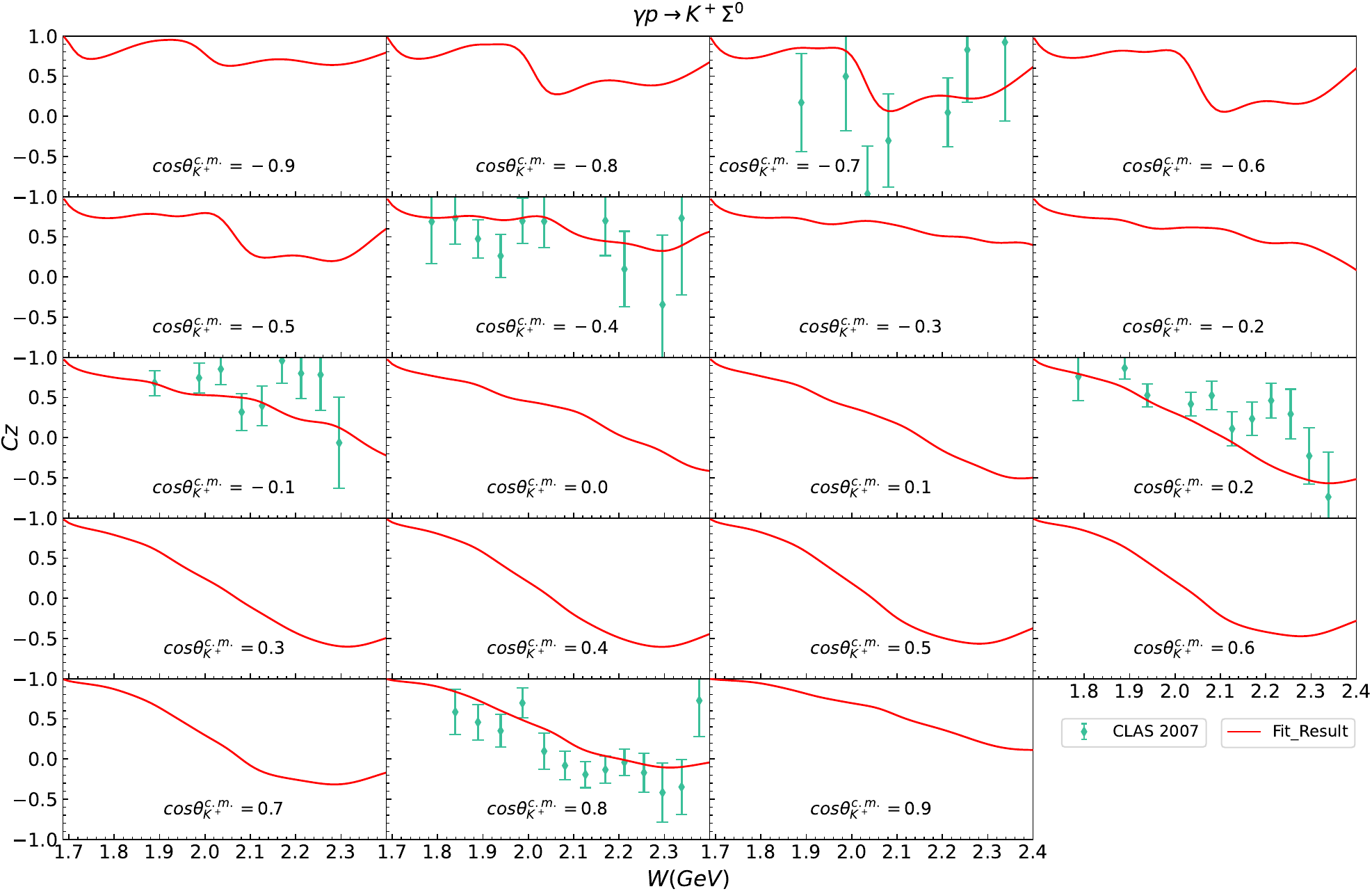}
    \caption{Beam-recoil double polarization $C_z$ for $\gamma p \to K^+ \Sigma^0$. Notation for the theoretical and experimental results is shown in the legend.}
    \label{Czp}
\end{figure}

\begin{figure}[]
    \includegraphics[scale=0.44]{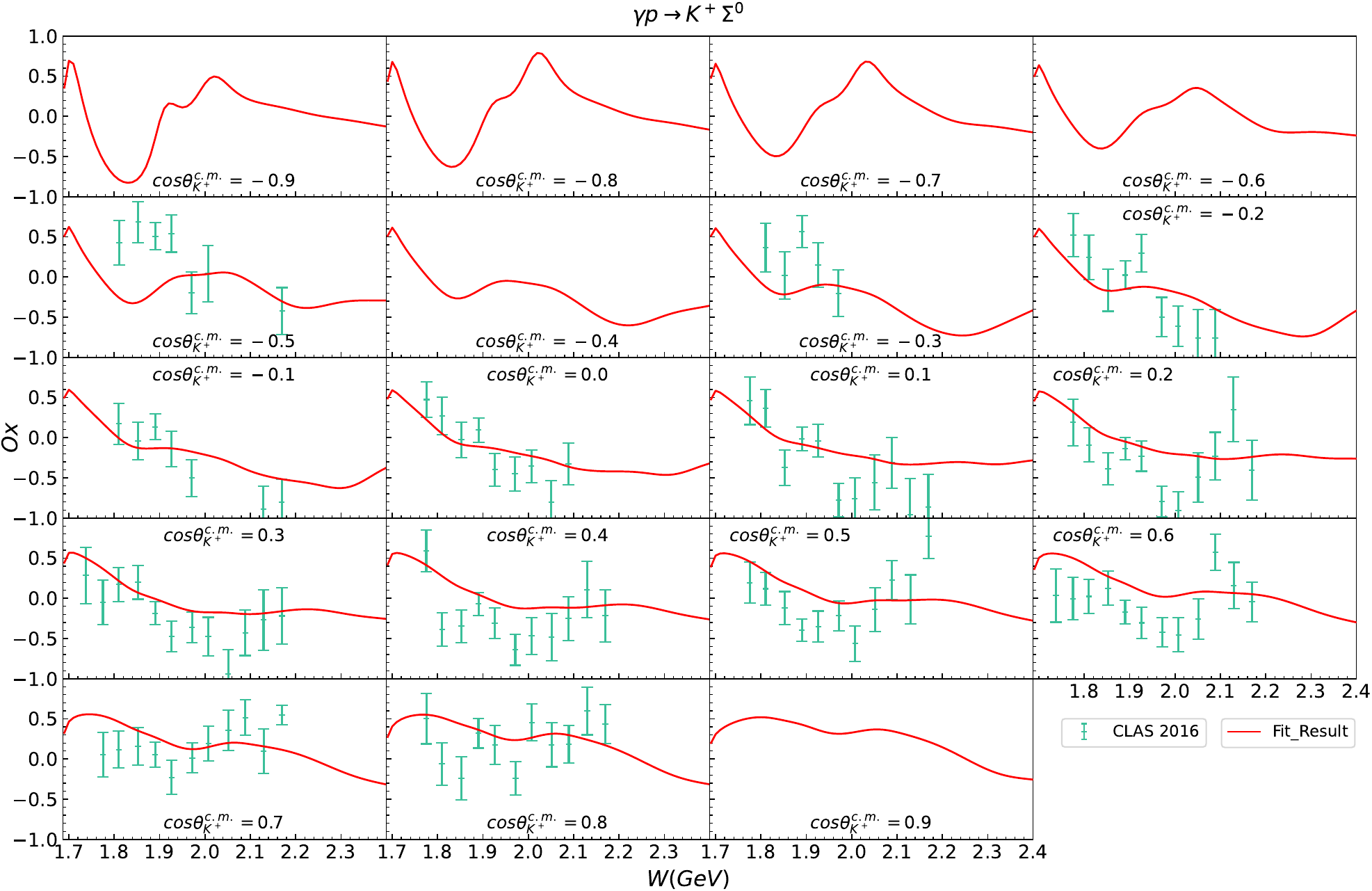}
    \caption{Beam-recoil double polarization $O_x$ for $\gamma p \to K^+ \Sigma^0$. Notation for the theoretical and experimental results is shown in the legend.}
    \label{Oxp}
\end{figure}

\begin{figure}[]
    \includegraphics[scale=0.44]{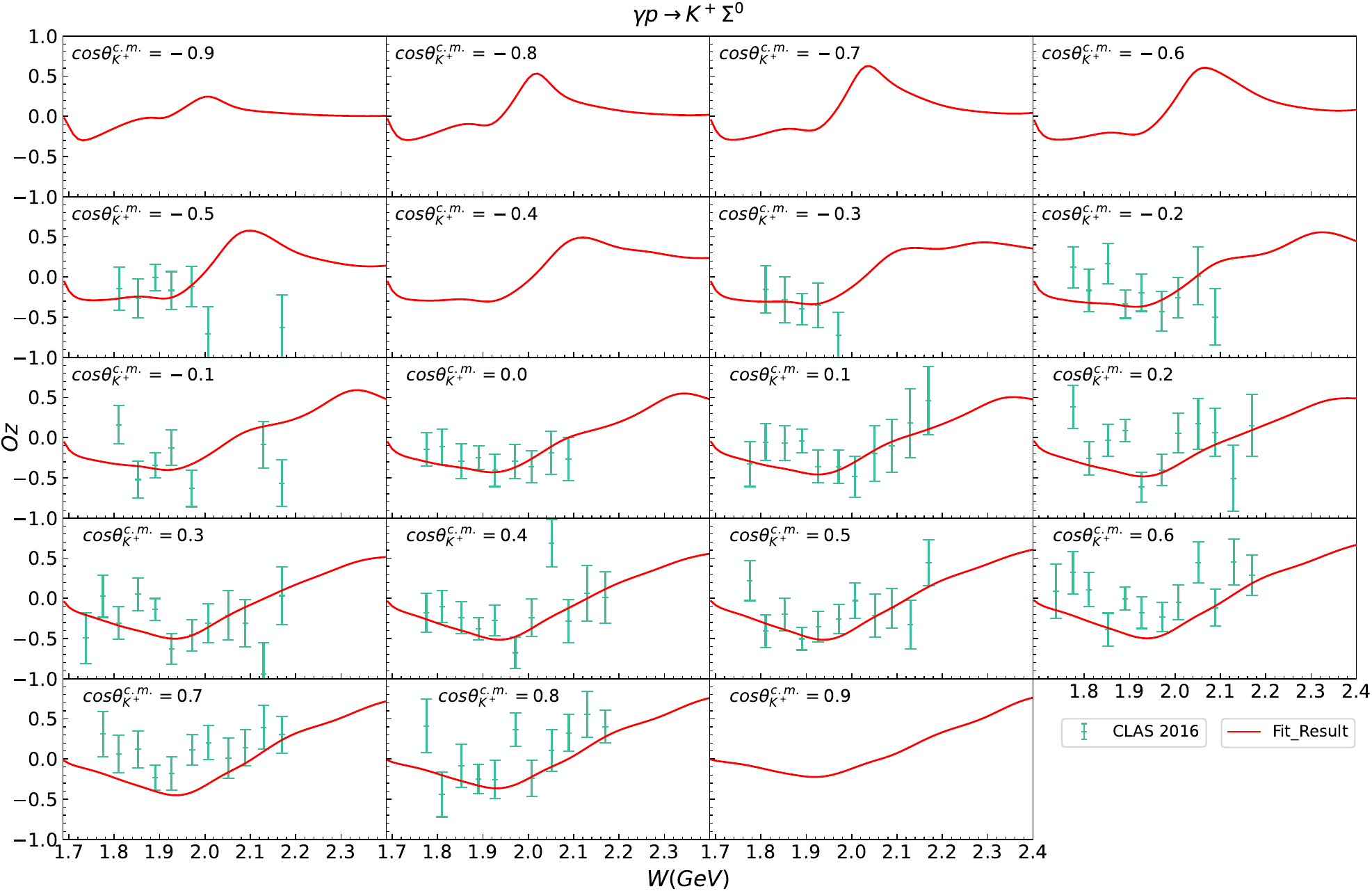}
    \caption{Beam-recoil double polarization $O_z$ for $\gamma p \to K^+ \Sigma^0$. Notation for the theoretical and experimental results is shown in the legend.}
    \label{Ozp}
\end{figure}

\begin{figure}[]
    \includegraphics[scale=0.44]{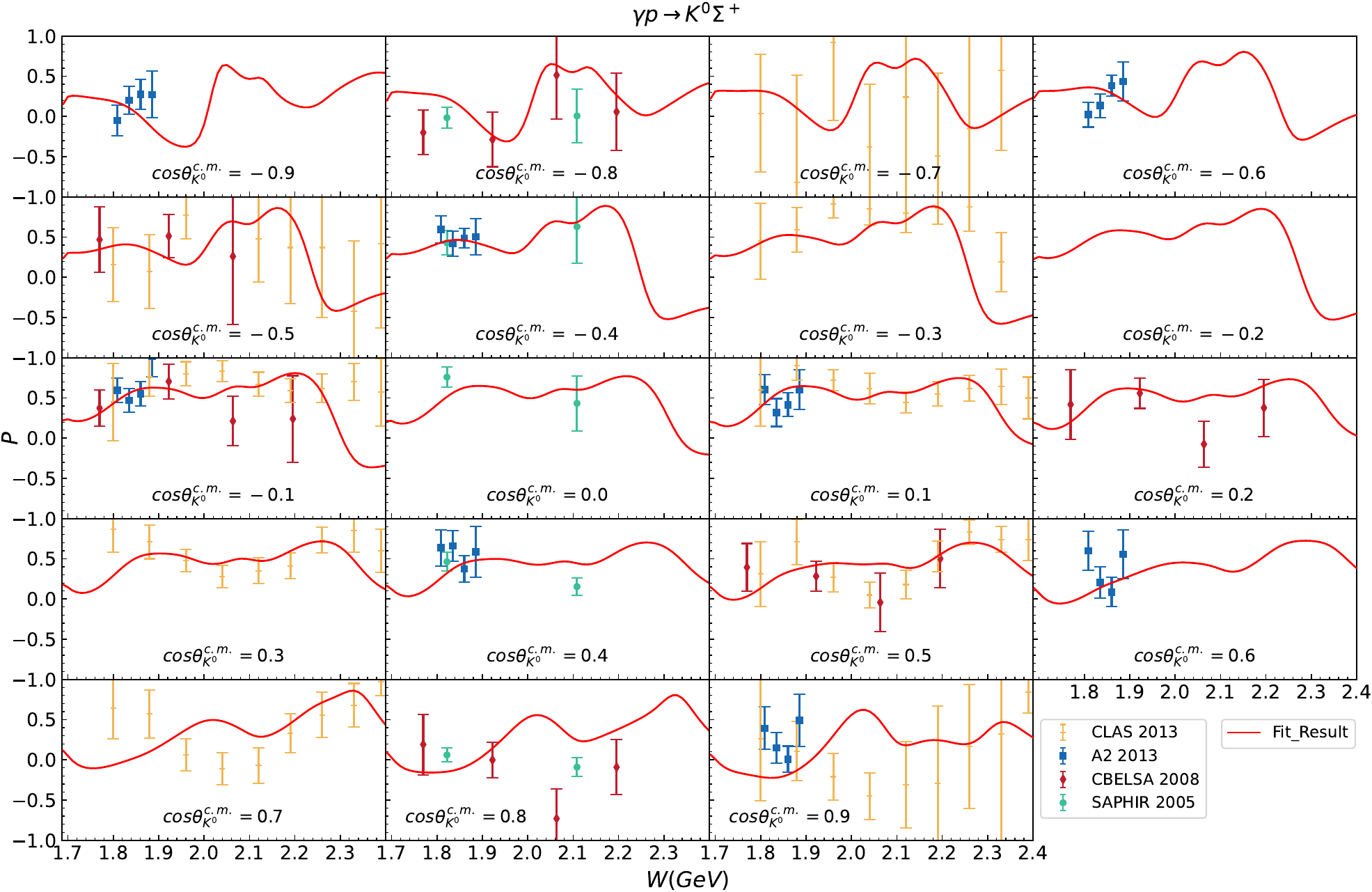}
    \caption{Recoil polarization $P$ for $\gamma p \to K^0 \Sigma^+$. Notation for the theoretical and experimental results is shown in the legend.}
    \label{P0}
\end{figure}

\begin{figure}[]
    \includegraphics[scale=0.44]{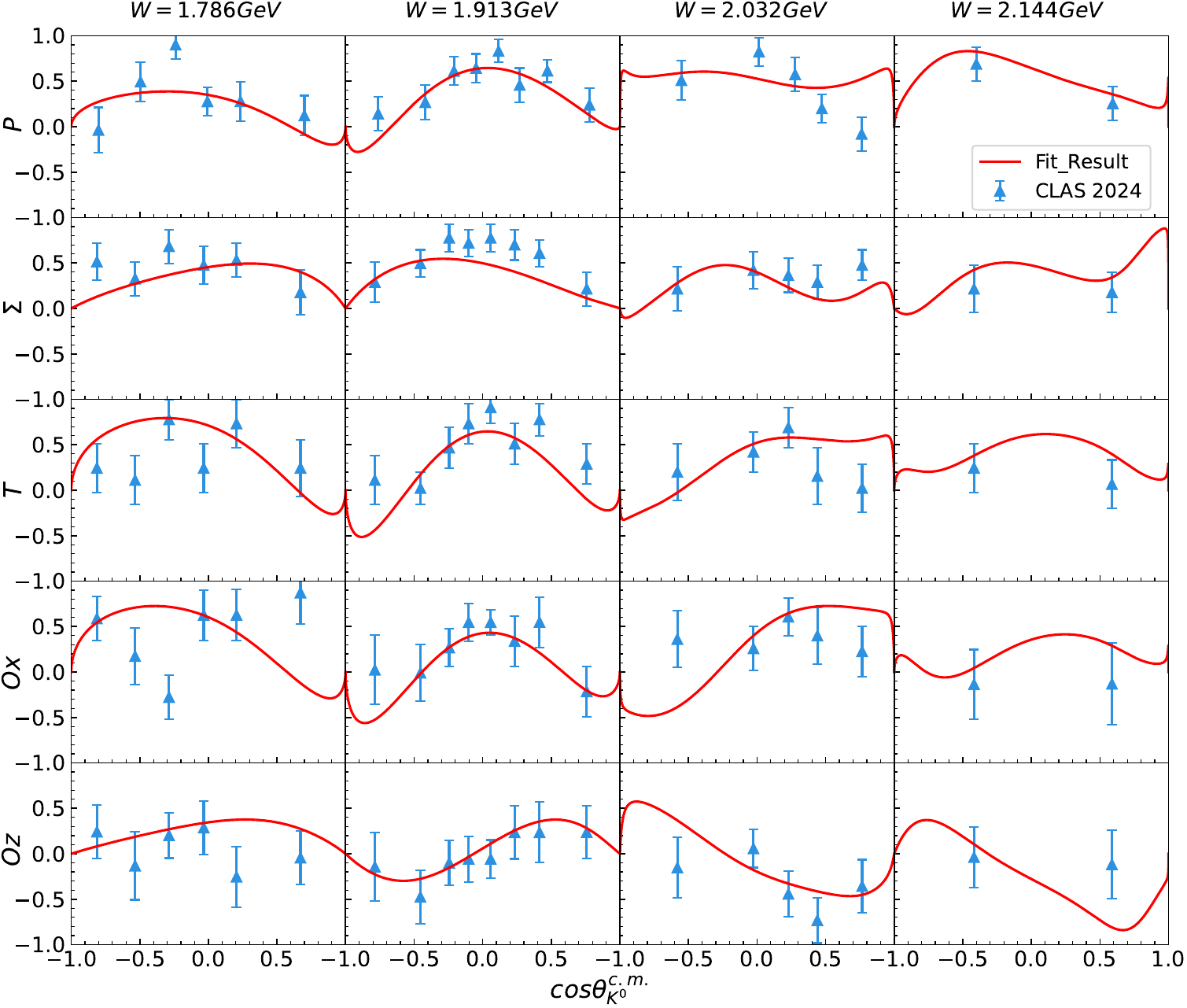}
    \caption{Recoil polarization $P$, Photon beam asymmetry $\Sigma$, Target asymmetry $T$, Beam-recoil double polarizations $O_x$ and $O_z$ as functions of the $cos \theta_{K^0}^{c.m.}$. All the experimental data are sourced from the CLAS 2024~\cite{CLAS:2024bzi}.}
    \label{CLAS2024}
\end{figure}

Figs.~\ref{Pp} to ~\ref{CLAS2024} display the polarization observables for $\gamma p \to K^+ \Sigma^0$ and $\gamma p \to K^0 \Sigma^+$ obtained in our theoretical calculations corresponding to the parameters listed in Tabs.~\ref{Table3} and ~\ref{Table4}. Almost all of the available experimental data shown in these figures can be well described, which is truly encouraging and demonstrates the effectiveness of our theoretical model. We also present predictions for some regions currently lacking experimental data, which can be compared with future experimental results. There are two points that need further explanation below.

First, as mentioned in Sec.~\ref{sec:Fitted results}, the $\chi^2/N_{data}$ of $\Sigma$ for $\gamma p \to K^+ \Sigma^0$ is relatively high, primarily due to the rather small error bars associated with the data points. In Fig.~\ref{Σp}, we can see that our theoretical results are in good agreement with the experimental data in most regions. However, due to the quite small error bars in the experimental data from CLAS 2016~\cite{CLAS:2016wrl}, even slight deviations can lead to a significant increase in the value of $\chi^2$. So we do not adjust the weight of it to improve its $\chi^2$. 

Second, the amount of experimental data for the polarization observables is still relatively limited, particularly for the reaction $\gamma p \to K^0 \Sigma^+$, and the precision of some available experimental data is also insufficient. 
These result in the experimental data still not being adequately constraining for our model parameters. We just provide a potential theoretical result based on the currently available experimental data for the reactions $\gamma p \to K^+ \Sigma^0$ and $\gamma p \to K^0 \Sigma^+$. However, more abundant and high-precision experimental data, particularly for the reaction $\gamma p \to K^0 \Sigma^+$, are necessary to further strengthen the constraints on the theoretical models.

\section{SUMMARY AND CONCLUSION}\label{sec:SUMMARY AND CONCLUSION}
Our previous studies revealed evidence of the strange molecular partners of $P_c$ states, $N(2080)3/2^-$ and $N(2270)3/2^-$, in the $\gamma p \to K^{*+} \Sigma^0 / K^{*0} \Sigma^+$ and $\gamma p \to \phi p$ reactions~\cite{Ben:2023uev,Wu:2023ywu}. Inspired by the experimental data of differential cross-sections for $\gamma p \to K^+ \Sigma^0$ from CLAS 2010~\cite{CLAS:2010aen}, which reveal some bump structures around $W$ = 1875, 2080 and 2270 MeV—corresponding to the Breit-Wigner masses of $N(1875)3/2^-$, $N(2080)1/2^- \&\ 3/2^-$, $N(2270)1/2^- , 3/2^- \&\ 5/2^-$—we decided to extend our previous work by investigating the effects of these six molecules, along with $N(1535)1/2^-$, as strange partners of $P_c$ molecular states in the reactions $\gamma p \to K^+ \Sigma^0$ and $\gamma p \to K^0 \Sigma^+$. Our theoretical model is based on an effective Lagrangian approach in the tree-level Born approximation~\cite{Wei:2022nqp,Wei:2023gvh}, and contains the contributions from $s$-channel with exchanges of $N$, $\Delta$, $N^*$(including the hadronic molecules with hidden strangeness), and $\Delta^*$; $t$-channel; $u$-channel; and the generalized contact term. Through some simplification settings in Sec.~\ref{sec:Fit parameters}, there are a total of 77 free parameters in the model listed in Tabs.~\ref{Table3} and ~\ref{Table4}, which represents a relatively reduced number of fit parameters. We then construct $\chi^{2}_{weight}$ through our theoretical model, incorporating nearly all available experimental data and associated weights listed in Tab.~\ref{Table1}. The fitted values of the free parameters are determined by minimizing the $\chi^{2}_{weight}$ with MINUIT.

The theoretical results based on the final fitted parameters in Tabs.~\ref{Table3} and ~\ref{Table4} are in good agreement with all available experimental data (cross-sections and polarization observables) for the $\gamma p \to K^+ \Sigma^0$ and $\gamma p \to K^0 \Sigma^+$ reactions, as directly indicated by the low $\chi^2 / N_{data}$ values in Tab.~\ref{Table2}. 
Achieving this is challenging for coupled-channel fits that involve two different reactions and data sets from various measurements, demonstrating the effectiveness of our theoretical model. 

In the results of cross-sections, contributions from the $s$-channel $\Delta^*$ resonance exchanges are more substantial for the reaction $\gamma p \to K^+ \Sigma^0$. In contrast, for the reaction $\gamma p \to K^0 \Sigma^+$, the contributions from $N^*$ and $\Delta^*$ resonance exchanges become comparable. This difference is attributed to the variation in the isospin factor $\tau$. Regarding the background, $s$-channel proton exchange, $t$-channel $K$ exchange, $u$-channel $\Sigma$ exchange and the interaction current have little contribution. The ground state $\Delta$ exchange has a relatively significant contribution, similar to other $\Delta^*$ resonances. In the reaction $\gamma p \to K^+ \Sigma^0$, the $t$-channel $K^* (892)$ and $K_1 (1270)$ exchanges provide considerable contributions of differential cross-sections at the forward angles in the high energy regions. Conversely, for the reaction $\gamma p \to K^0 \Sigma^+$, the contribution from $K_1(1270)$ exchange becomes negligible, while the contribution from $K^*(892)$ exchange increases a lot.

As for the molecules, the $N(1875)3/2^-$ exchange provides the largest contributions among molecules. Alongside $N(1535)1/2^-$, exchanges of these two molecules contribute across a wide energy range due to their relatively large widths. Together with contributions from $s$-channel general resonance exchanges, they help construct the overall structure of the cross-sections, particularly the peak at $W \approx 1900$ MeV. Notably, the substantial interference effects between the contributions from $s$-channel general resonance exchanges and molecule exchanges, are one of the important reasons for the significant differences in the magnitudes of the cross-sections for $\gamma p \to K^+ \Sigma^0$ and $\gamma p \to K^0 \Sigma^+$. In addition, $N(2080)1/2^- \&\ 3/2^-$ and $N(2270)1/2^- , 3/2^- \&\ 5/2^-$ exchanges are mainly responsible for the peak structures around $W$ = 2080 and 2270 MeV, respectively. And the contributions from these molecules with different spins are roughly comparable, showing no obvious preference for any particular spin. Moreover, compared with the HFF-P3 model in Ref.~\cite{Clymton:2021wof}, our results of total cross-sections exhibit distinct peaks around $W$ = 2080 and 2270 MeV, indicating the significant effects of molecules.

For the results of polarization observables, all experimental data are well described. The predictions for some regions currently lacking experimental data are also presented, which can be compared with future experimental results. However, the amount of experimental data for the polarization observables is still relatively limited, particularly for the reaction $\gamma p \to K^0 \Sigma^+$. Meanwhile, the cross-section data for $\gamma p \to K^0 \Sigma^+$ are also much sparser compared to those for $\gamma p \to K^+ \Sigma^0$. These result in the experimental data still not being adequately constraining for our model parameters.

Furthermore, to verify the stability of these fitted results with the molecular masses fixed, we use this set of fitted values as initial values to perform the fitting with the molecular masses above the $K\Sigma$ threshold released. The convergent fitted values for most of the molecular masses fall within a variation range of 100 MeV. Notably, the mass of $N(2270)3/2^-$ remains almost unchanged, indicating a particularly strong tendency for $N(2270)3/2^-$ to contribute in this region.

More abundant experiments are necessary to further strengthen the constraints on the theoretical models, particularly for the reaction $\gamma p \to K^0 \Sigma^+$, due to the effects of isospin factors and the unbalanced datasets. Hopefully further experiments can distinguish various models.

\section{ACKNOWLEDGMENTS}\label{sec:ACKNOWLEDGMENTS}
We thank Feng-Kun Guo, Jia-Jun Wu, Shu-Ming Wu, Zhen-Hua Zhang and Chao Tang for their useful discussions
and valuable comments. We acknowledge the HPC Cluster of the Institute of Theoretical Physics, Chinese Academy of Sciences (ITP-CAS) for computational support. This work is supported by the
NSFC and the Deutsche Forschungsgemeinschaft (DFG,
German Research Foundation) through the funds provided
to the Sino-German Collaborative Research Center
TRR110 “Symmetries and the Emergence of Structure in
QCD” (NSFC Grant No. 12070131001, DFG Project-ID
196253076-TRR 110), by the NSFC Grant
No. 12047503, and by the Chinese Academy of Sciences
(CAS) under Grant No. XDB34030000.

\bibliography{Documents.bib} 

\end{document}